\documentclass[twocolumn, twocolappendix]{aastex631}
\usepackage{natbib}
\usepackage{wrapfig}
\usepackage{amsmath}
\usepackage{hyperref}
\usepackage{enumerate}
\begin{document}

\title{Improved Tomographic Binning of 3x2pt Lens Samples: Neural Network Classifiers and Optimal Bin Assignments}

\correspondingauthor{Irene Moskowitz}
\email{iwm15@physics.rutgers.edu}

\author[0000-0002-2206-8589]{Irene Moskowitz}
\affiliation{Department of Physics and Astronomy, Rutgers, The State University of New Jersey, 136 Frelinghuysen Rd, Piscataway, NJ 08854, USA}

\author[0000-0003-1530-8713]{Eric Gawiser}
\affiliation{Department of Physics and Astronomy, Rutgers, The State University of New Jersey, 136 Frelinghuysen Rd, Piscataway, NJ 08854, USA}

\author[0000-0002-9964-1005]{Abby Bault}
\affiliation{Department of Physics and Astronomy, University of California, Irvine, CA 92697, USA}

\author[0000-0002-7767-5044]{Adam Broussard}
\affiliation{Department of Physics and Astronomy, Rutgers, The State University of New Jersey, 136 Frelinghuysen Rd, Piscataway, NJ 08854, USA}

\author[0000-0001-8684-2222]{Jeffrey A. Newman}
\affiliation{Department of Physics and Astronomy and PITT PACC, University of Pittsburgh, Pittsburgh, PA 15260, USA}

\author[0000-0001-9789-9646]{Joe Zuntz}
\affiliation{Institute for Astronomy, Royal Observatory Edinburgh, University of Edinburgh, Edinburgh EH9 3HJ, United Kingdom}

\author{The LSST Dark Energy Science Collaboration}

\begin{abstract}
Large imaging surveys, such as LSST, rely on photometric redshifts and tomographic binning for 3x2pt analyses that combine galaxy clustering and weak lensing. In this paper, we propose a method for optimizing the tomographic binning choice for the lens sample of galaxies. We divide the CosmoDC2 and Buzzard simulated galaxy catalogs into a training set and an application set, where the training set is non-representative in a realistic way, and then  estimate photometric redshifts for the application sets. The galaxies are sorted into redshift bins covering equal intervals of redshift or comoving distance, or with an equal number of galaxies in each bin, and we consider a generalized extension of these approaches. We find that bins of equal comoving distance produce the highest dark energy figure of merit of the initial binning choices, but that the choice of bin edges can be further optimized. We then train a neural network classifier to identify galaxies that are either highly likely to have accurate photometric redshift estimates, or highly likely to be sorted into the correct redshift bin. The neural network classifier is used to remove poor redshift estimates from the sample, and the results are compared to the case when none of the sample is removed. We find that the NNCs are able to improve the figure of merit by $\sim 13\%$, and are able to recover $\sim 25\%$ of the loss in the figure of merit that occurs when a non-representative training sample is used. 
    
\end{abstract}

\section{Introduction}

The $\Lambda$CDM paradigm has been highly successful in explaining the evolution and accelerated expansion of the universe, with approximately $30\%$ of the energy density of the universe present in visible and dark matter, and the remaining $70\%$ in dark energy \citep{Planck2018}. The nature of this dark energy is not currently well-understood, but the simplest model is a cosmological constant, $\Lambda$, in which the energy density of the dark energy does not change as the universe expands. A proposed alternative, the \textit{w}CDM model, allows the energy density to evolve with time. The nature of the dark energy has been investigated by Stage-III dark energy experiments, including the Dark Energy Survey (DES, \citealt{DES}), the Hyper Suprime-Cam Subaru Strategic Program (HSC, \citealt{hsc_overview}), the Kilo-Degree Survey (KiDS, \citealt{kids}) and the Extended Baryonic Oscillation Spectroscopic Survey (eBOSS, \citealt{eboss}). Stage-IV surveys have begun \citep{desi}, with the Vera C. Rubin Observatory's Legacy Survey of Space and Time (LSST, \citealt{ivezic2010}), Euclid \citep{euclid}, and Roman \citep{roman}, scheduled to come online in the next few years. 

Two powerful cosmological probes are weak lensing (see, for example, \citealt{WeakLensingCFHTLenS, WeakLensingRCSLenS, WeakLensingHSC, WeakLensingKiDS, WeakLensingDES}) and large scale structure (see, for example, \citealt{PressSchecterClusters, ClustersDavis, ClustersDesjacques, ClustersHaloModels, ClustersSDSS, ClustersDES}). Two-point auto-correlation functions are computed between the shears of background (source) galaxies (shear-shear correlations), the locations of foreground (lens) galaxies (galaxy clustering), and the cross-correlation between the shears of the source galaxies and the locations of the lens galaxies (galaxy-galaxy lensing). This combination of three types of two-point correlation functions is called the 3x2 point (3x2pt) method, and has been used previously by DES \citep{DES} and KiDS \citep{kids}.

If very accurate distances (i.e. spectroscopic redshifts) could be obtained for all the galaxies in a survey, the two-point correlation functions could be computed in 3D. However, many Stage-III and -IV surveys, including LSST, are large imaging surveys, and it is not possible to obtain spectroscopic redshifts for most of the galaxies that will be observed. Data for the 3x2pt method is therefore analyzed tomographically, with galaxies sorted into bins of estimated redshift. After sorting, the exact redshift of each galaxy is no longer important, as angular two-point correlations are computed within and between each bin and only the redshift distribution of each bin is needed to interpret the results.

One way to sort galaxies into redshift bins is to estimate their photometric redshifts (photo-$z$'s), which are less precise than spectroscopic redshifts, but can be obtained for a much larger sample of galaxies without prohibitive observing time requirements. There are two main categories of photo-$z$ estimation codes: template-fitting and machine learning. Template-fitting codes use either observed or theoretical galaxy template spectra and apply filter transmission curves to predict the photometry for a galaxy at a given redshift. The observed photometry is then matched to this library to find the best fit combination of template and redshift (e.g., \citealt{eazy, template_photo_z, hyperz}). In contrast, machine learning codes use a variety of techniques, including self-organized maps (e.g., \citealt{SOM_Masters, SOM_wright}), neural networks (e.g., \citealt{annz, netz}), and random forests (e.g., \citealt{Carrasco_Kind_2013}), to associate input colors and magnitudes with a redshift estimate. A method for improving photo-$z$ estimates using machine learning post-processing in the form of a Neural Network Classifier (NNC) was recently introduced by \citet{Broussard}. 

Methods for optimizing the tomographic binning for the source sample in LSST 
have been addressed
in \citet{tomo_challenge}. In this paper, we present a method for optimizing the tomographic binning for the {\em lens} sample. Where the methods for optimizing the source sample in \citet{tomo_challenge} are restricted to \textit{riz} photometry due to the use of the metacalibration \citep{metacal1, metacal2} technique for computing shears, the lens sample is under no such restriction, and we are free to use all the photometric information available for optimization.
 We make use of data simulated to resemble expected LSST observations, and use a realistically non-representative training sample of galaxies with spectroscopic redshifts. 
Next, we investigate extensions of three different tomographic binning methods to identify the optimal combination.
We then use two variations of the NNC developed by \citet{Broussard} to select galaxies with the most accurate photo-$z$ estimates and calculate the resulting dark energy Figure of Merit (FoM) of the galaxy clustering signal. 

In section 2, we present the simulated galaxy catalogs used in this work, which have been split into training, validation, and application samples. In section 3, we describe how the non-representative training sample is built, and introduce the photo-$z$ estimation pipeline, along with two different ways to train the NNC used for selecting the final application sample; the different choices of binning; and how the FoM of each binning choice is evaluated. Section 4 presents the results, and we conclude in section 5.

\section{Simulated Data}

We make use of two simulated galaxy catalogues, CosmoDC2 \citep{cosmoDC2} and Buzzard \citep{buzzard}. 

\subsection{CosmoDC2}
The CosmoDC2 simulation covers $440$ deg$^2$ on the sky, and is complete to a magnitude $r=28$ and contains galaxies up to a redshift of 3. 
The CosmoDC2 catalog was created using dark matter particles from the Outer Rim simulations \citep{outer_rim}. Galaxies with a limited set of specified properties were then assigned to dark matter halos using the UniverseMachine simulations \citep{universemachine} and the GalSampler technique \citep{galsampler}. These objects were matched to the outputs of the Galacticus model \citep{galacticus}, which generated complete galaxy properties. 

\subsection{Buzzard}
The Buzzard catalog was developed to simulate DES Year 1 data, and contains galaxies up to a redshift of 2.3 over an 1120 deg$^2$ area  and complete to $r\sim 26$. The Buzzard catalog is based on dark matter simulations from L-GADGET2 \citep{gadget2}. Galaxies were added to dark matter halos based on abundance matching using \textsc{AddGals} \citep{addgals}. Galaxies are drawn from a distribution of luminosities and overdensities, and matched to halos of the same overdensity. SEDs are assigned to match the SED-luminosity-density relationship measured from SDSS data. 

\subsection{The DESC Tomographic Challenge}
The LSST Dark Energy Science Collaboration (DESC) Tomographic Challenge \citep{tomo_challenge} was designed as a first step towards optimizing the tomographic binning for the source sample of galaxies in a 3x2pt analysis. Participants in the challenge were invited to use any method they liked for determining the binning, but were limited to using (\textit{g})\textit{riz} photometry. This choice was made to reflect the fact that the DESC plans to compute shears using the metacalibration technique \citep{metacal1, metacal2}, which requires selections to be performed only in bands in which the PSF is well measured.

For use in the Tomographic Challenge, the CosmoDC2 and Buzzard catalogs 
had noise added to simulate real LSST Year 1 observations using the DESC \textsc{TXPipe} \footnote{\href{https://github.com/LSSTDESC/TXPipe}{https://github.com/LSSTDESC/TXPipe}} framework and following the methodology of \citet{ivezic2010}.  The catalogs 
were further cut based on signal to noise and galaxy size to simulate the selection criteria used for real lensing catalogues. Galaxies in the full simulated catalogues were kept for the final sample of galaxies if the combined \textit{riz} signal to noise satisfied $S/N > 10$ and the size satisfied $T/T_{\rm psf} > 0.5$. Here, \textit{T} measures the squared, deconvolved radius of the galaxy, and is the trace $I_{xx}+I_{yy}$ of the moments matrix, and $T_{\rm psf}$ is derived from the assumed full width at half maximum (FWHM) of the Rubin PSF,  $T_{\rm fwhm}=0.75$ arcseconds. After these cuts, the CosmoDC sample contained 36 million objects, and the Buzzard sample contained 20 million objects. From these samples,  $25\%$ were randomly assigned to the training sample, $50\%$ to the validation sample, and the remaining $25\%$ to the application sample, making the training sample fully representative. For this work, we have made use of the Tomographic Challenge training samples, which we have split into training and validation samples, and the Tomographic Challenge validation samples as our application sample.

\section{Methods}
The general process for optimizing the tomographic binning method consists of six steps: 

\begin{enumerate}[i]
    \item To simulate realistic training conditions, use Hyper Suprime Cam (HSC) photometric and spectroscopic galaxies to create a realistically non-representative training sample. 
    \item Use the training sample to train a method for photometric redshift estimation. In the work, we use Trees for photo-$z$ (TPZ, \citet{Carrasco_Kind_2013}). This training is then applied to estimate photometric redshifts for the application sample. 
    \item Select bin edges based on photo-$z$ estimates for the sample using extensions upon three simple approaches.
    \item Train a Neural Network Classifier (NNC) to obtain a confidence in the accuracy of the photo-$z$ estimates from TPZ.
    \item Sort galaxies into the chosen bins both with and without a confidence cut based on the NNC.
    \item Compute the FoM of the resulting bins for comparison with methods produced as part of the DESC Tomographic Challenge \citep{tomo_challenge}.
\end{enumerate}

The following subsections describe this process in detail.   

\subsection{Non-representative Training Sample}
While the Tomographic Challenge methods were built and run using a training sample that was fully representative of the application sample, this will not be the case with LSST data. Previous attempts to account for the effects of a non-representative training sample can be found in \citet{Beck_2017} and \citet{gpz_nonrep}.

To divide the CosmoDC2 and Buzzard data sets into training and application samples that are non-representative, we make use of the second data release (HSC DR2) of the Hyper Suprime Cam Subaru Strategic Program (HSC SSP) \citep{hsc}. From the HSC Wide (300 deg$^2$ at \textit{i} $<26.2$) \citep{hsc_dr2} catalog, we selected objects with \textit{i} $\le 25$, and either \textit{g} $\le 27.5$, \textit{r} $\le 27.7$ or \textit{z} $\le 26.2$, which mimics early LSST observations, as noted in \citet{Broussard}. After this photometric cut, there are 3,469,800 remaining objects; a color-magnitude diagram of a randomly selected subset of these objects is shown in the middle panel of Figure 1 of \citet{Broussard}. This sample is named HSC phot. 

In addition, some galaxies in the HSC Wide field have been matched to spectroscopic catalogs from zCOSMOS \citep{zcosmos}, WiggleZ DR1 \citep{wigglez}, DEEP3 \citep{deep3_2011}, PRIMUS DR1 \citep{primus1, primus2}, DEEP2 DR4 \citep{deep2}, UDSz \citep{udsz1, udsz2}, VVDS \citep{vvds}, 3D-HST \citep{3dhst1, 3dhst2}, VIPERS PDR1 \citep{vipers}, FMOS-COSMOS \citep{fmos_cosmos1, fmos_cosmos2}, GAMA DR2 \citep{gama}, SDSS DR12 \citep{ssds_dr12}, and the SDSS IV QSO catalog \citep{ssds_qso}. There are 307,193 objects meeting the same photometric requirements as above that also have matched spectroscopic redshifts. A color-magnitude diagram of this sample, named HSC spec is shown in \citet{Broussard} in the left panel of their Figure 1.

\begin{figure*}
\plotone{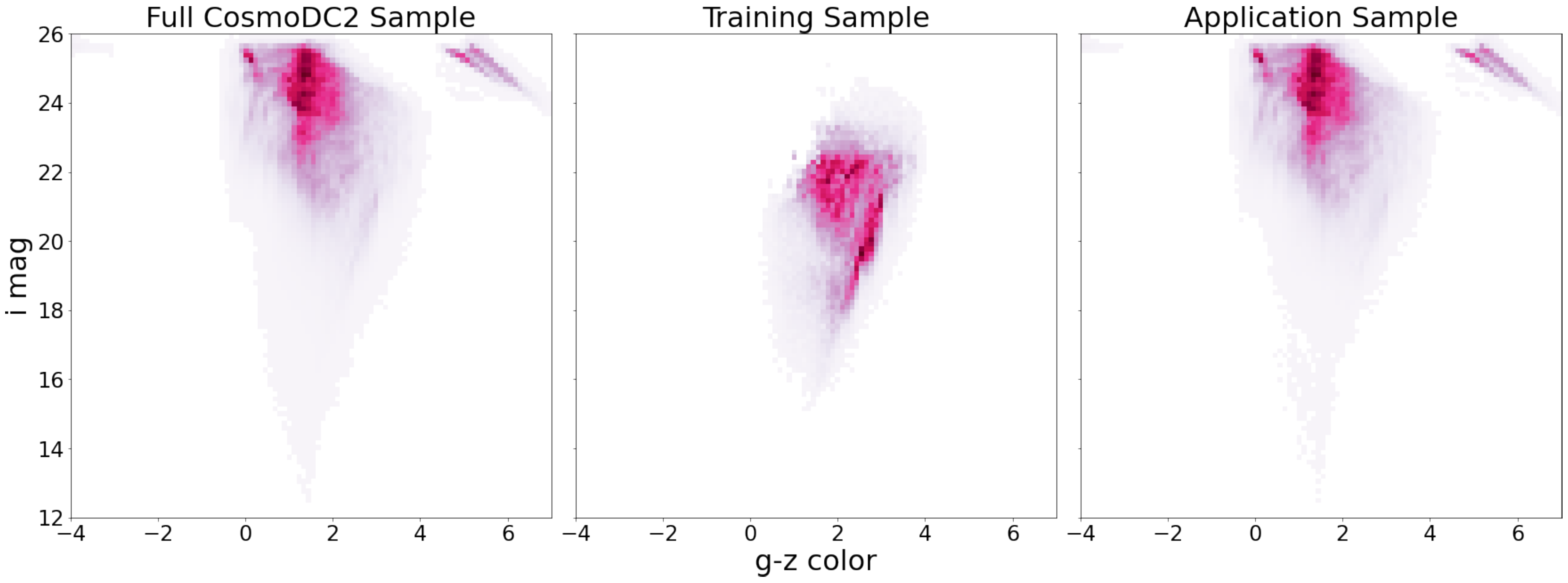}
\caption{Partitioning the original CosmoDC2 sample into non-representative training and application samples. 
The left panel shows the original color-magnitude distribution of the catalog, while the middle and right panels show the selected training and application samples, respectively. The color bar  shows the relative density per pixel within each panel on an arbitrary scale. \label{fig:DC2_unrep}}
\end{figure*}

To build the non-representative training sample, we divide the space of \textit{i}-band magnitude and (\textit{g-z}) color  into 100x100 bins. In each bin, we compute the ratio between HSC spec objects with spectroscopic redshifts and HSC phot objects.  Even at fixed brightness and color, galaxy spectral features can vary significantly, and spectroscopic confirmation rates are often significantly lower for higher-redshift galaxies in a given color-mag bin.  To model this, we find the 99th percentile in spectroscopic redshift within each bin, and assign that as $z_{\rm max}$ for that color-mag bin. 

Objects in the CosmoDC2 and Buzzard catalogs are then sorted into the same bins in color-magnitude space. In each bin, objects with photometric redshifts greater than $z_{\rm max}$ are automatically assigned to the application sample.  The remaining objects are then randomly assigned to the training or application sample based on the ratio of HSC spec to HSC phot objects. If there were no HSC spec objects in a given color-mag bin, all of the objects were assigned to the application sample. This method of producing a non-representative training sample has been incorporated into the DESC RAIL\footnote{\href{https://github.com/LSSTDESC/RAIL}{https://github.com/LSSTDESC/RAIL}} framework as the GridSelection degrader. Inevitably, this is still an approximation of true non-representativeness, which can get worse than what is presented here. As an example, a spectroscopic survey might use $B$-band selection, which is not directly replicable with the $ugrizy$ photometry available for LSST. The non-representative sample presented here may still prove to be somewhat optimistic, unless efforts are made to obtain spectroscopic samples with a greater coverage of color-mag-redshift space.

\begin{figure}
\plotone{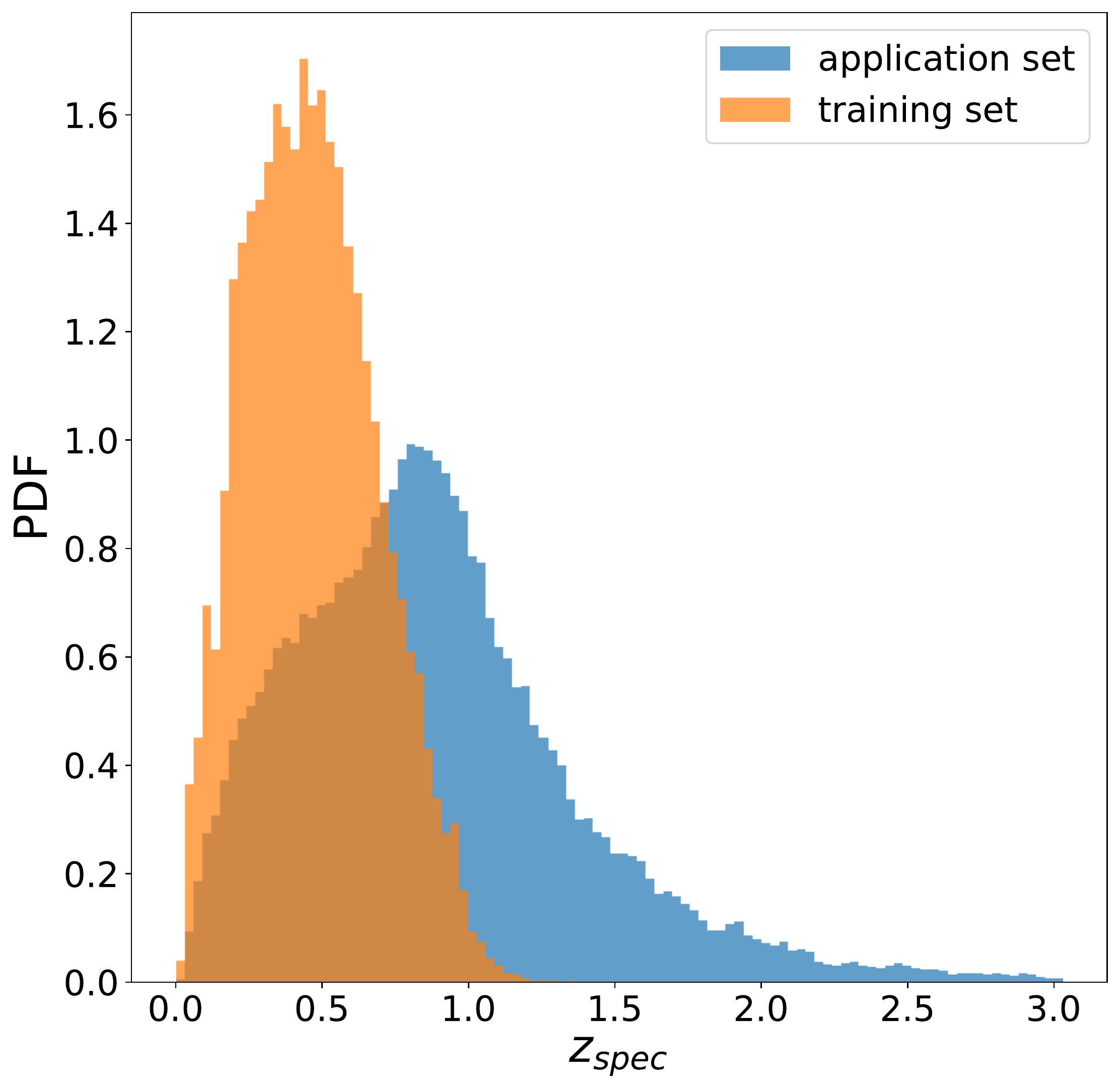}
\caption{Normalized spectroscopic redshift distributions for the whole CosmoDC2 sample (blue) and the selected non-representative training sample (orange) Note that the heights of the peaks are not comparable; the overall size of the training sample is much smaller than the application sample.
\label{fig:unrep_redshift}}
\end{figure}

Figure \ref{fig:DC2_unrep} shows the partitioning of the CosmoDC2 sample into training and application samples. Critically, the training sample is brighter (the training sample has median $i$-band magnitude of 21.1) than the application sample (median $i$-band magnitude of 23.9), as is typical for spectroscopic follow-up of deep photometric surveys. The mean color of the training sample is also redder, as is expected. The median ($g-z$) color of the training sample is 2.20, while the median color of the application sample is 1.55. The training sample shown in Figure \ref{fig:DC2_unrep} is used for training TPZ and the NNCs, and represents a spectroscopic sample, while the application sample is used for estimating photo-zs, NNC confidences, and the binning results, and represents LSST data.

A further illustration of the non-representative nature of the training sample we have built can be seen in Figure \ref{fig:unrep_redshift}, which shows the normalized true redshift distribution of the selected application sample compared to that of the selected training sample.  The redshift distribution of the training sample falls off above a redshift of 1.0 with a tail to $z=1.5$, while the application sample falls off more gradually and extends out to a redshift of 3. In the application sample, $34\%$ of the objects are at $z>1.0$, while in the training sample, only $0.085\%$ of objects are at $z>1.0$. Additionally, the peak of the redshift distribution of the training sample occurs at lower redshifts than for the application sample. Note that the height of the peaks in Figure \ref{fig:unrep_redshift} are not directly comparable; the training sample is much smaller overall.  

\subsubsection{Training Sample Size}
We also use the HSC DR2 to construct a realistically sized training sample. There are 3,469,800 objects in the HSC DR2 catalog that meet the photometric cuts described in section 3.1, and 307,193 objects with spectroscopic redshifts. This translates to a training sample that is $\sim8\%$ of the application sample. After splitting the Tomographic Challenge data sets into our non-representative training and application samples, we select approximately 3.4M for the application sample, and a training sample that's $\sim 7.6\%$ the size of the application sample, a smaller fraction that the original Tomographic Challenge sample. This overall training sample is split into equally sized training and validation samples for TPZ. The TPZ validation sample is then further split into $2/3$ to be used for the NNC training sample, and $1/3$ used for the NNC validation sample. 

\subsection{Photometric Redshifts from TPZ}
TPZ is a machine learning method of photo-$z$ estimation that uses random forests, and we used the regression mode for this work. We used 100 trees for the random forest, with a minimum leaf size of 30. TPZ also allows us to choose $m_*$, the number of features that are randomly selected for splitting a given tree node; we provide TPZ with 11 features for training, and chose $m_*=6$. 

We provide the following 11 features as inputs to TPZ for training: the apparent magnitudes in the \textit{ugrizy} filters and the colors (\textit{u-g}), (\textit{g-r}), (\textit{r-i}), (\textit{i-z}) and (\textit{z-y}). Galaxies with a non-detection in a given band are assigned a magnitude of 30 in that band. We additionally tried adding the band triples, defined as the difference between neighboring colors in \citet{Broussard}. In analogy to the manner in which colors act as a derivative of the magnitudes, band triples act as a second derivative. From the set of features we provide, the available band triples are [(\textit{u-g}) - (\textit{g-r})], [(\textit{g-r}) - (\textit{r-i})], [(\textit{r-i}) - (\textit{i-z})], and [(\textit{i-z}) - (\textit{z-y})]. We find here that including the triples in the training does improve the photo-$z$ estimation by a marginal amount, but the Neural Network Classifiers do just as well without them, and it is too computationally expensive to be worth including them.  

TPZ provides three outputs for each galaxy in the application sample: $z_{\rm phot}$, the mean of the PDF; $\sigma_{\rm TPZ}$, the associated Gaussian uncertainty in the $z_{\rm phot}$ estimate; and $z_{\rm conf}$, the integrated probability between $z_{\rm phot} \pm \sigma_{\rm TPZ}\left(1+z_{\rm phot}\right)$. 

\subsection{Figure of Merit Calculation}
 
Tomographic Challenge methods were evaluated using two metrics: the signal-to-noise ratio (SNR) of the angular power spectra derived from the $n_i(z)$ distribution of each bin \textit{i}, and the figure of merit (FoM). The FoM comes in two flavors: the $S_8-\Omega_m$ FoM, and the $w_0 - w_a$ FoM, which constrains the dark energy equation of state parameters \citep{detf}.  We choose to focus on the $w_0 - w_a$ FoM (hereafter simply FoM) as our metric for optimizing the binning choices since this is specific to the DESC goal of measuring the dark energy equation of state parameters. The binning optimization procedure we outline in the rest of the paper can be used for other science goals, but a different metric may be preferable in those cases.

Each metric has contributions from galaxy clustering and weak lensing components, which are combined to form the 3x2pt FoM. Since assumptions made about galaxy bias in the Tomographic Challenge lead to the total 3x2pt FoM being dominated by the clustering signal, we continue to use the 3x2pt FoM despite focusing on the choice of lens sample. In doing so, we have assumed a source sample that is equivalent to the lens sample, even though the source sample cannot be selected using all of the bands in this work. This makes the exact FoM values reported here somewhat optimistic, but it is the relative changes in the FoM between binning choices that are most important. In order to directly compare the tomographic binning method described here to the results of the Tomographic Challenge, we use the same method for calculating the FoM.  

The FoM is calculated using the inverse area of a Fisher matrix ellipse representing the area of the posterior PDF of the two parameters we want to measure. The FoM is then related to the constraining power of the chosen binning scheme: a higher FoM corresponds to a smaller area, meaning tighter constraints on $w_0$ and $w_a$. The Fisher matrix is given by:

\begin{equation}
    F = \left(\frac{\partial \mu}{\partial \theta}\right)^T C^{-1} \left(\frac{\partial \mu}{\partial \theta}\right),
\end{equation}
where $\mu$ is the vector of theoretical predictions for the $C_{\ell}$ spectra for each pair of tomographic bins, $\theta$ are the list of cosmological parameters included in the theory, and \textit{C} is a Gaussian estimate of the covariance between them following \citet{takada_and_jain}. To compute the $C_{\ell}$, we use the true redshift distribution, $n_i(z)$, which is the distribution of the actual redshifts of the galaxies in each bin, rather than the distribution of photometric redshifts. The $C_{\ell}$ are calculated for 100 values of $\ell$ between 100 and 2000 More details can be found in the appendix of \citet{tomo_challenge}. 

The FoM is then calculated as:

\begin{equation}
    \textrm{FoM} = \frac{1}{2\pi \sqrt{\textrm{det}\left(\left[F^{-1}\right]_{p_1,p_2}\right)}},
\end{equation}
where $p_1=w_0$ and $p_2=w_a$ extracts the 2x2 submatrix of $F^{-1}$ corresponding to the dark energy equation of state parameters. The list of parameters used for calculating the Fisher matrix is the native combination in the DESC core cosmology library \citep{ccl}: $\Omega_c$, $\Omega_b$, $H_0$, $\sigma_8$, $n_s$, $w_0$ and $w_a$. The Tomographic Challenge FoM calculation does not include any galaxy bias values as nuisance parameters, which leads to the overall 3x2pt FoM being dominated by the galaxy clustering signal.

Our analysis required some changes to the FoM calculation from the Tomographic Challenge. The Challenge used a hard coded density of galaxies on the sky, which we updated to reflect the actual galaxy density of the CosmoDC2 and Buzzard simulations. The CosmoDC2 simulation covers 440 deg$^2$ on the sky, and the final CosmoDC2 sample contained 36M objects, so this density was used as the starting density. Although we did not use the entire Challenge catalog, we used the appropriate sky density for the sample to obtain comparable FoM values. Instead of hard coding a sky density, this ensures that sample cuts using the NNCs will become susceptible to shot noise in bins with few galaxies, thereby reducing the FoM as appropriate. We also updated the hard-coded fraction of the sky covered by the survey. We implemented two options: scaling the survey area to match the fraction of the catalog we used to maintain a comparable survey density, or assuming the sky density for a given simulation but scaled to the LSST year 10 survey area. The rest of this paper reports results for the assumed year 10 area. 

\subsection{Selecting Bin Edges}
\begin{figure*}[ht!]
\plotone{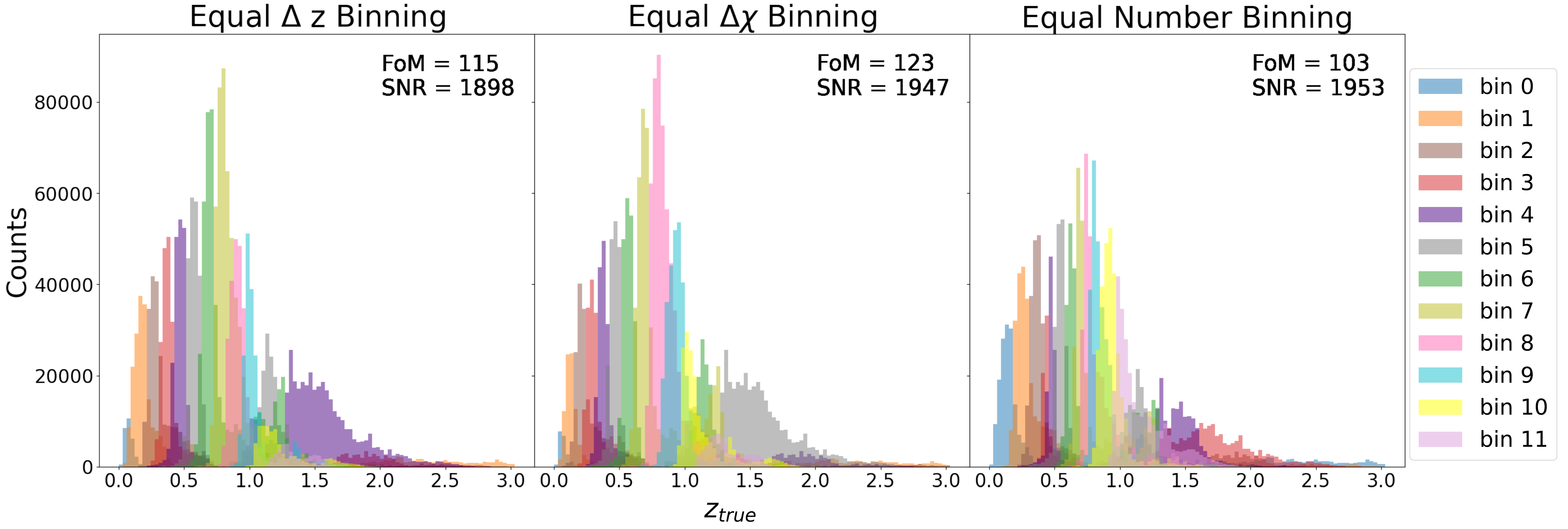}
\caption{Comparisons of three base binning methods using the CosmoDC2 sample. Left panel: equal $\Delta$z binning. Middle panel: equal $\Delta \chi$ binning. Right panel: equal number binning. Equal $\Delta \chi$ binning achieves the highest $w_0 - w_a$ figure of merit}.    \label{fig:binning_methods}
\end{figure*}

Once photo-$z$ estimates are made, galaxies can be binned. Based upon the optimization of galaxy clustering SNR explored in \citet{Broussard}, we utilize 12 tomographic bins covering $0<z<3$. We start with three possible binning options: bin edges equally spaced in redshift (equal $\Delta z$ binning), bin edges equally spaced in comoving distance (equal $\Delta \chi$ binning), and bins with equal numbers of galaxies in each bin (equal number binning). Examples of the three binning methods, along with the computed FoM values, are shown for the CosmoDC2 sample in Figure \ref{fig:binning_methods}. Note that for binning choices related to distance, like equal $\Delta z$ and equal $\Delta \chi$, this results in a couple of sparsely populated bins at the high redshift tail of the sample. In these cases, there are $\sim10$ effective bins up to $z\sim1.2$ that contain most of the information. This is not the case for equal number binning.

 Figure \ref{fig:binning_methods} illustrates that the equal $\Delta \chi$ binning case produces the highest figure of merit value, while equal number binning performs the worst out of our three base binning choices. Equal $\Delta \chi$ binning has not been explored much in the literature, but the Tomographic Challenge method that achieved the highest FoM also utilized equal $\Delta \chi$ bins \citep{tomo_challenge}. Although they did not include equal $\Delta \chi$ bins in their analysis, \citet{binning_euclid} also found that equally spaced redshift bins performed better than equal number bins for galaxy clustering and galaxy-galaxy lensing when using 13 bins, and the two performed about evenly with 10 bins, which is consistent with our findings. This is in contrast with \citet{marvin2021}, which found that equal number binning out-performed equal $\Delta z$ binning for a Euclid-like survey, although we find agreement with these results when the SNR is used as the evaluation metric. \citet{binning_shear} and \citet{binning_soms} found that bins equally spaced in redshift performed better than bins with equal numbers of galaxies for cosmic shear analysis, although neither were for the same number of bins used here. \citet{binning_shear} concluded that while having a large number of bins equally spaced in redshift provides the best constraints on cosmological parameters, bins with equal numbers of galaxies were better for $N_{bin}<20$. Meanwhile, \citet{binning_soms} only tested on 3, 4, and 5 tomographic bins.  

In addition to the three common binning choices above, there are many more ways that the bin edges could be chosen, so we introduce a new parameter, $\mathcal{M}$, that allows us to investigate an  extended family of possible bin definitions. We define $\mathcal{M}$ as 

\begin{equation}
    \mathcal{M} = \int_0^{z_{\rm max}}\left(\frac{dN}{dz}\right)^{\alpha} \left(\frac{d\chi}{dz}\right)^{\beta} dz, 
    \label{eq:optimal_binning}
\end{equation}
where $\alpha$ and $\beta$ will be chosen to maximize the final signal-to-noise of the resulting bins. Once $\mathcal{M}$ is calculated for a given $\alpha$ and $\beta$, $\mathcal{M}$ is divided into equal intervals based on the number of bins desired, in this case 12, then those values of $\mathcal{M}$ are converted back into redshift values for the bin edges by interpolation.

There are three sets of $\alpha$ and $\beta$ values that can recover the previous binning choices. When $\alpha = \beta = 0$, $\mathcal{M} = z_{\rm max}$, and we recover equal $\Delta z$ binning. When $\alpha = 0$ and $\beta = 1$, $\mathcal{M} = \chi_{max}$, and we recover equal $\Delta \chi$ binning. When $\alpha=1$ and $\beta = 0$, $\mathcal{M} = N$, the total number of galaxies in the sample, and we recover equal number binning. Other values for $\alpha$ and $\beta$ 
represent variations on these approaches. 

\subsection{The ``Outlier'' Neural Network Classifier}
Since LSST is expected to observe enough galaxies that it will not be shot-noise limited, it should be possible to improve the FoM of the bins shown in Figure \ref{fig:binning_methods}, or determined by Equation \ref{eq:optimal_binning}, by removing galaxies with poor photo-$z$ estimates. After training and applying TPZ, we implement the Neural Network Classifier (NNC) as described in \citet{Broussard}, which we refer to here as the ``Outlier'' NNC.   

The NNC takes as its inputs the three outputs from TPZ, $z_{\rm phot}$, $\sigma_{\rm TPZ}$ and $z_{\rm conf}$, along with the size of the galaxies, the \textit{i}-band magnitude and the same colors that were used for training TPZ. The NNC consists of four fully connected hidden layers, with [100, 200, 100, 50] neurons using a Scaled Exponential Linear Unit (SELU) activation function. The output neuron produces a value between 0 and 1, indicating redshift accuracy confidence, by using a sigmoid function with a binary cross-entropy loss function.  An NNC confidence output close to 1 indicates high confidence in the accuracy of the photo-$z$ estimate, while a confidence output near 0 indicates a probable outlier. The output NNC confidence values are then used to make a sample cut.

To define what counts as an accurate photo-$z$ estimate when training the NNC, 
we must choose a maximum acceptable value for the ``accuracy parameter'' $\left|\Delta z \right|/\left(1+z\right)$. 
\citet{Broussard} tested accuracy parameters between 0.04 and 0.15. They found that while the value of the accuracy parameter has an impact on the fit, if the goal is to cut a certain fraction of the sample e.g., the worst $10\%$, the value of the accuracy parameter does not change which galaxies are in that worst $10\%$, only what the threshold confidence value is.  In other words, we are free to choose the accuracy parameter to obtain reasonable looking confidence values, since the accuracy parameter changes the confidence, but not the ordering. We chose to use $\left|\Delta z \right|/\left(1+z\right) < 0.07$.

\begin{figure}[h!]
\plotone{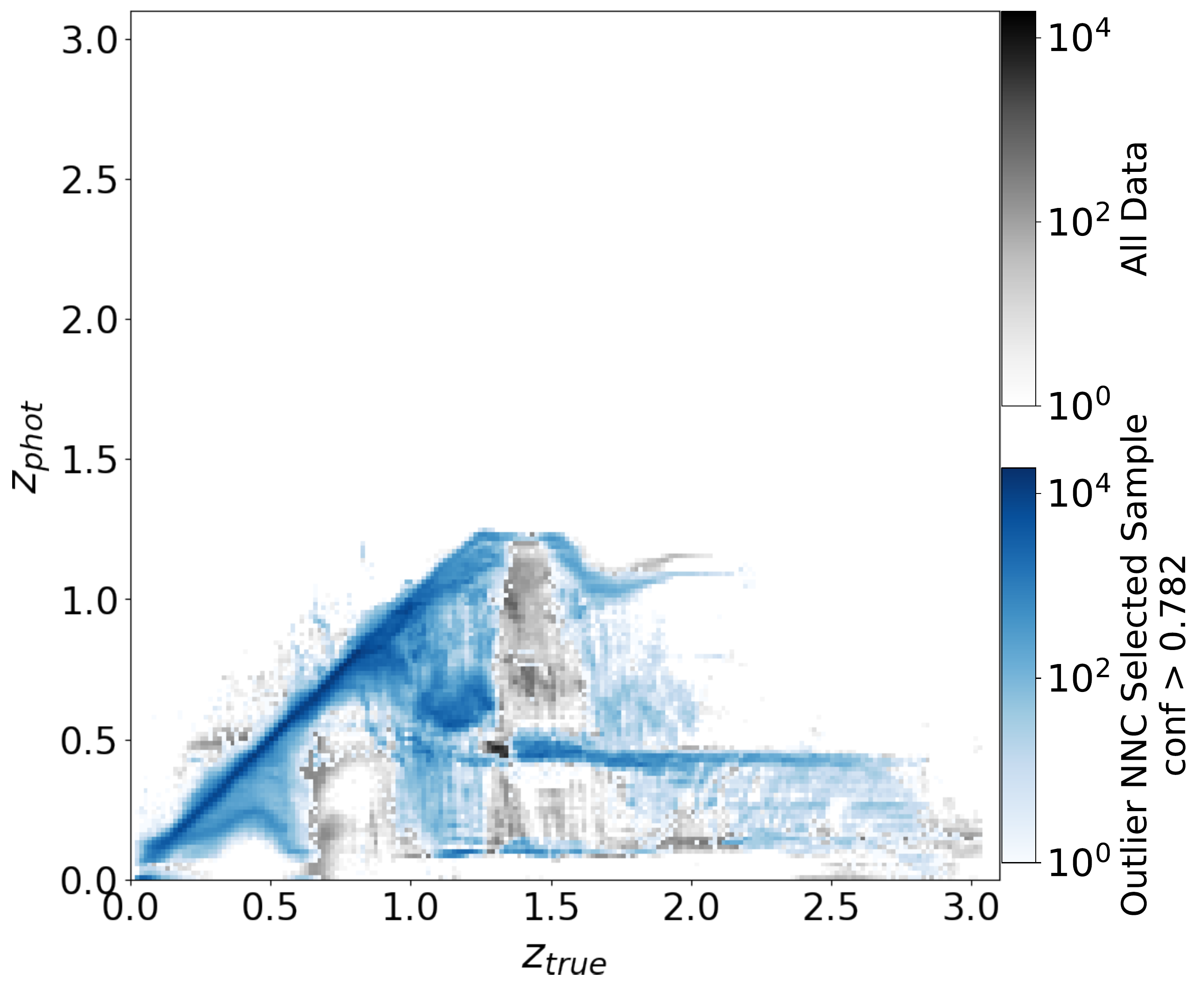}
\caption{photo-$z$ estimate vs. true redshift for the entire CosmoDC2 sample (black) and the Outlier NNC selected sample (blue). The NNC confidence cut has been chosen to retain $75\%$ of the original sample.\label{fig:outlier_sample}}
\end{figure}

Figure \ref{fig:outlier_sample} shows the results of implementing a confidence cut based on the NNC confidence levels for the CosmoDC2 sample. The confidence cut of 0.782 was chosen to retain $75\%$ of the original sample, as this will be shown later to produce the highest FoM with this NNC. Since TPZ was trained on a non-representative training sample with a maximum redshift lower than the maximum redshift of the application sample, the high redshift portion of the application consists entirely of outlier photo-$z$ estimates. The NNC successfully excludes a large portion of these major photo-$z$ outliers, reducing the outlier fractions (defined as $\Delta z/(1+z) > 0.15$) by about a third from $f_{\rm out} = 0.30$ for the full sample to $f_{\rm out} = 0.20$ for the NNC selected sample. The Outlier NNC selects galaxies close to the one-to-one line representing accurate photo-$z$ estimates when trained on a representative sample.

We can also see in Figure \ref{fig:outlier_sample}, and later in Figure \ref{fig:misclass_num_sample}, that TPZ produces concentrations near $z_{phot}\sim0.1$ and $z_{phot}\sim0.5$. We would expect bands arising from degeneracies in galaxy colors to form diagonal bands, not horizontal, so this is likely not an effect of degeneracies in galaxy SEDs at different redshifts. It appears that TPZ has found some preferred redshifts. Investigation of the cause and mitigation of this effect is left for future work.

\subsection{The ``Misclassification'' NNC}

One of the benefits of tomographic binning is that once a galaxy has been sorted into a bin, its individual redshift value is no longer important; angular correlation functions are computed for the population of galaxies within the bin. In this sense, it doesn't matter if the photo-$z$ estimate is highly accurate as long as it's good enough for most galaxies to be sorted into the correct tomographic bin. Following this logic, we implement a new method for training a ``Misclassification'' NNC.  Instead of training it to assign low confidence values to galaxies likely to be outliers, we trained this NNC to assign low confidence values to galaxies likely to be sorted into the incorrect bin. 

The Misclassification NNC has a more complicated training routine than the outlier NNC. With the Outlier NNC, we are free to chose the bin edges after applying the NNC and making the confidence cuts.  However, 
since the definition of the correct bin will depend on the choice of bin edges,
we must now chose the bin edges before training the NNC. Once photo-$z$ estimates are determined for the application sample, we use the entire sample to define the bin edges, using either equal $\Delta z$, equal $\Delta \chi$, equal number binning, or the optimized binning determined by equation \ref{eq:optimal_binning}. Then objects in the NNC training sample are sorted into those bins by their estimated photo-$z$, and we determine if the galaxy was sorted into the correct bin by comparing to the bin its true redshift would have placed it in. Galaxies in the correct bin are assigned an accuracy value of 1, while galaxies placed in any incorrect bin are assigned an accuracy value of 0. 

We also train a separate Misclassification NNC for each bin, which is then applied only to objects sorted into that bin. As an example, a `bin 3 NNC' is trained on objects in the training sample that have been sorted into bin 3, and is applied only to objects in the application sample that are sorted into bin 3. Instead of identifying objects across the whole sample that are likely to be misclassified, this version of the Misclassification NNC identifies galaxies that are likely to be incorrectly sorted into each bin. This is an attempt to make it easier for the NNC to learn the photometric features associated with each bin, since the photometry associated with being sorted into the wrong bin are not necessarily consistent over all bins. This version of the Misclassification NNC performs better than the version trained on the entire sample, and is the version we use going forward.

\begin{figure}[t]
\plotone{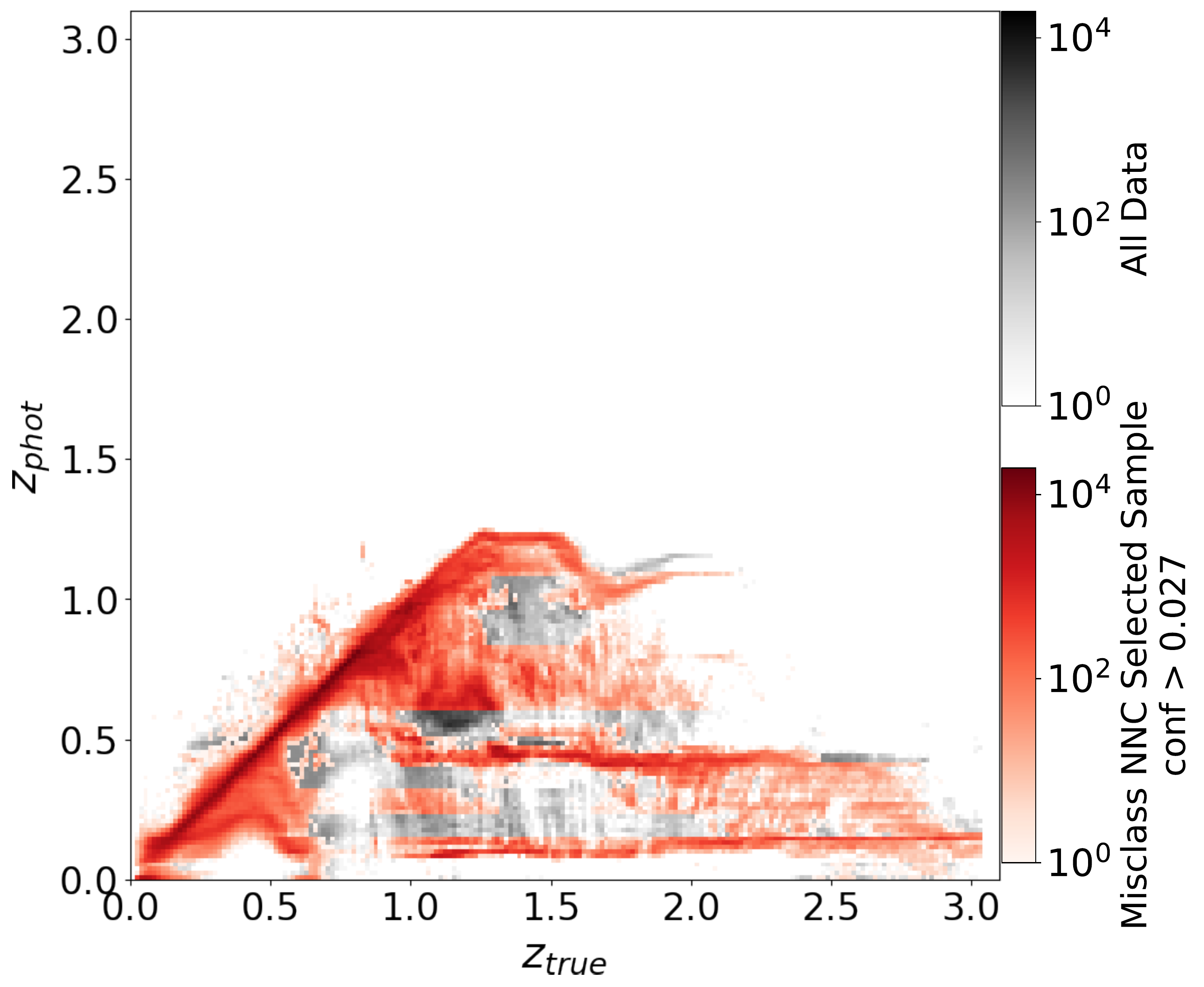}
\caption{The sample selected by the Misclassification NNC (red) trained on 12 equal number bins. The  confidence cut was selected to retain $75\%$ of the sample, as in Figure \ref{fig:outlier_sample}. Rectangular features are caused by the Misclassification NNC only paying attention to shifts between bins rather than redshift errors within a single bin.   \label{fig:misclass_num_sample}}
\end{figure}

Figure \ref{fig:misclass_num_sample} shows the sample selection using the Misclassification NNC. The NNC was trained on 12 equal $\Delta \chi$ bins, since this was the best performing of the base binning choices. The confidence cut was selected to retain $75\%$ of the original sample, as in Figure \ref{fig:outlier_sample}. The sample selected by the Misclassification NNC looks quite different to the sample selected by the Outlier NNC. Boxy horizontal features are caused by the NNC only caring if the photo-$z$ places the galaxy into the correct bin, and correspond to the bin edges. In particular, in the equal $\Delta \chi$ binning case, the bins get progressively wider as the redshift increases, the highest redshift bin is quite wide, since there are not as many galaxies at those redshifts, and a modestly larger fraction of the outliers in the upper right of Figure \ref{fig:misclass_num_sample} are kept in the sample compared to the Outlier NNC selection;  this is consistent with the Misclassification NNC only caring about objects that shuffle between bins, as the highest-redshift bin is broad.  

\section{Results}
We first determine the optimal choice of bin edges and the optimal fraction of the sample to retain with the NNCs.  
After the sample selections are made with the NNCs, the remaining galaxies are sorted into the optimized bin edges, which are determined using the redshift distribution for the full sample. We compare the optimized binning choice to the previous best binning choice, equal number binning.  Results in this section are shown for the CosmoDC2 simulation, and we leave the results for Buzzard to Appendix \ref{buzzard_appendix}, except where we discuss differences between the two simulations.

\subsection{Optimal Binning Approaches}

We calculate the FoM resulting from the bins produced by various combinations of $\alpha$ and $\beta$ in equation \ref{eq:optimal_binning}. In order to do these calculations, we must make a choice for the redshift distribution, $n(z)$ (equivalent to $\frac{dN}{dz}$ in equation \ref{eq:optimal_binning}). We have done these calculations for three versions of $n(z)$: the photo-$z$ $n(z)$, made up of the point estimates of the photo-$z$ for each galaxy, a version of the photo-$z$ $n(z)$ that has been smoothed with a Gaussian filter, or the true $n(z)$. In all three cases, we find slightly different combinations of $\alpha$ and $\beta$ produce the optimal binning choice, but they all find that equal number binning is not the optimal choice. We show results for the true $n(z)$ going forward.

\begin{figure}
\plotone{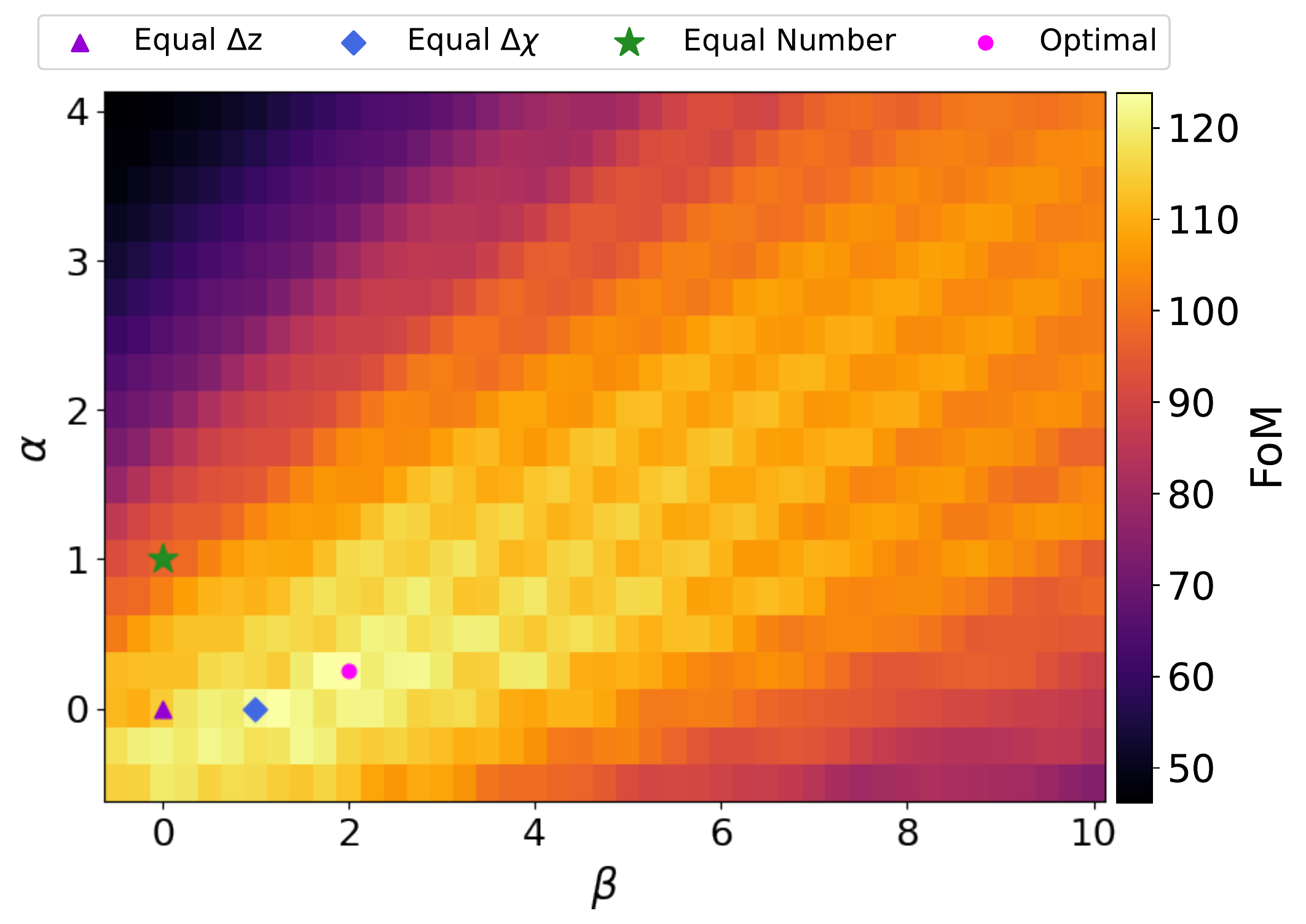}
\caption{The figure of merit of the resulting bins for different values of $\alpha$ and $\beta$. The maximum FoM is shown as a pink circle, while the previous binning choices of equal $\Delta$z, equal $\Delta \chi$ and equal number are shown as a purple triangle, blue diamond and green star, respectively. \label{fig:alpha_beta_plots}}
\end{figure}

Figure \ref{fig:alpha_beta_plots} shows the FoM values for a grid of different values of $\alpha$ and $\beta$ for CosmoDC2. The maximum FoM of 124 is achieved at $\alpha=0.25$ and $\beta=2.0$. We take this to be our optimized binning choice. The bins edges resulting from this choice of $\alpha$ and $\beta$ are listed in Table \ref{table:summary_table}. Buzzard reaches the maximum FoM at $\alpha$ and $\beta$ corresponding to equal $\Delta \chi$ binning, see Appendix \ref{buzzard_appendix}. 

\subsection{FoM vs. Retained Fraction}
Since LSST will not be shot noise limited, removing some of the galaxies with poor photo-$z$ estimates should improve the FoM. However, if too much of the sample is removed, we will reach the limit where shot noise increases sufficiently to drop the FoM below what it would be if we kept the entire sample.

\begin{figure}
\plotone{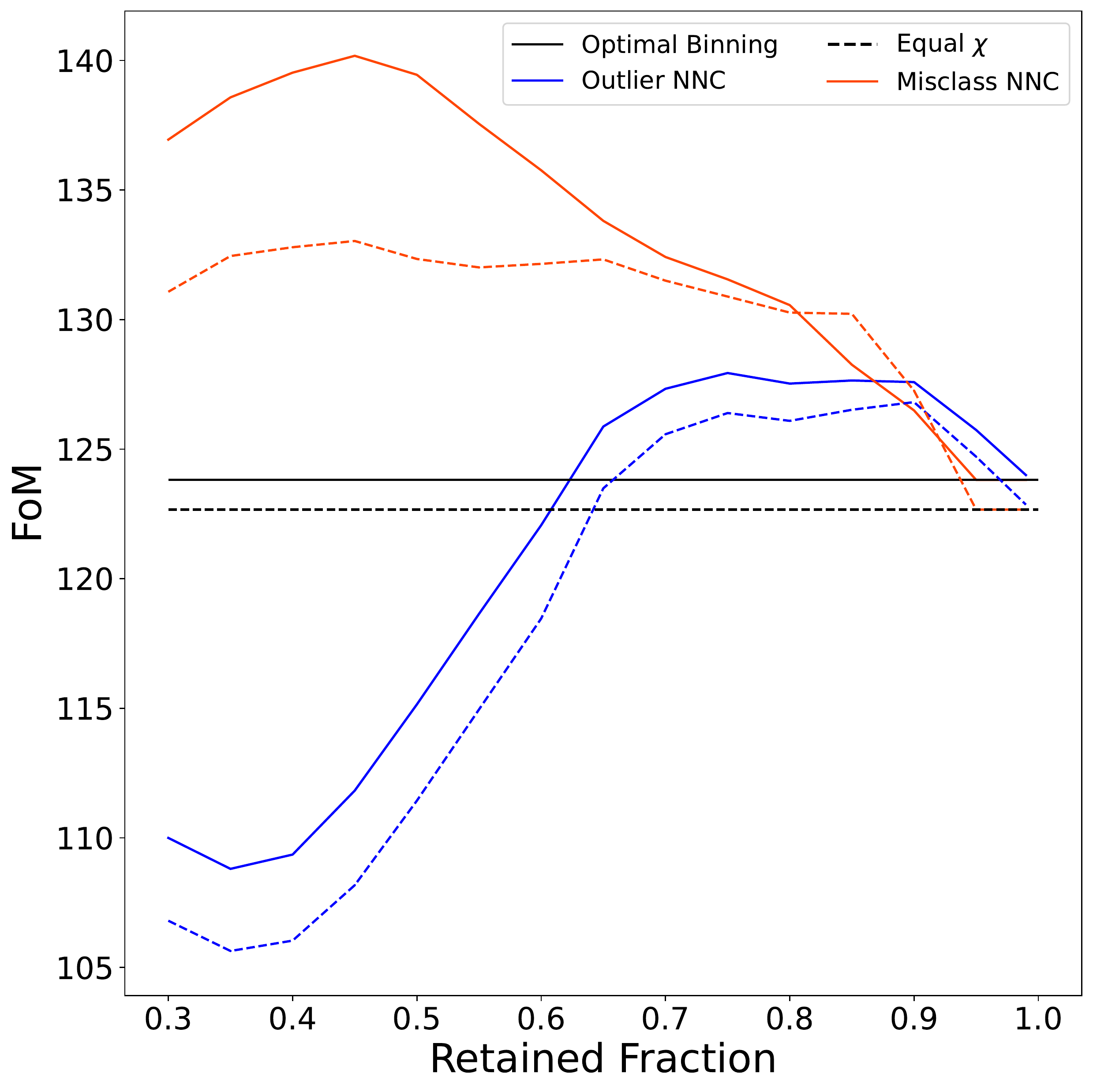}
\caption{The FoM as a function of the amount of the sample retained after cuts for CosmoDC2. Solid lines represent the FoM for the optimized binning choice with $\alpha=0.25$ and $\beta=2.0$. Dashed lines represent equal $\Delta \chi$ binning. Solid and dashed horizontal lines are the FoM for the full sample for the optimized bins and equal $\Delta \chi$ bins respectively. \label{fig:retained_frac_dc2}}
\end{figure}

\begin{figure}
\plotone{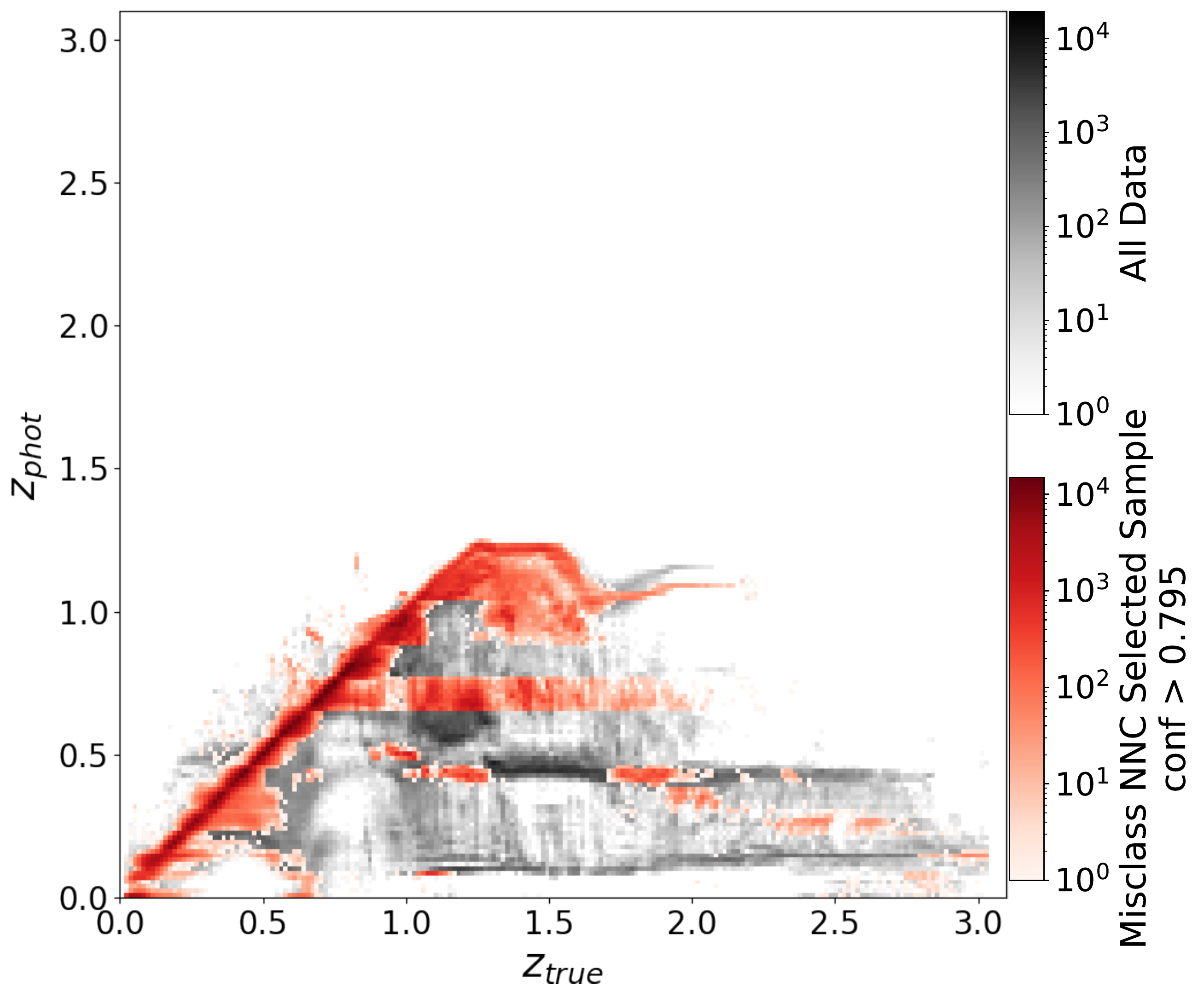}
\caption{The sample selected by the Misclassification NNC trained on the optimized bin choice, defined by $\alpha=0.25$ and $\beta=2.0$. Here we have retained $45\%$ of the original sample, which achieves the highest FoM for the Misclassification NNC. \label{fig:optimal_misclass_sample}}
\end{figure}

We test the range over which the NNC sample cuts will improve the FoM and find the optimal fraction of the sample to retain (the retained fraction). Figure \ref{fig:retained_frac_dc2} compares the performance of the FoM as a function of the retained fraction for each type of NNC and binning choice for CosmoDC2. The results for Buzzard are shown in Appendix \ref{buzzard_appendix}. It can be seen that the Outlier NNC performs better at slightly higher retained fractions, while the Misclassification NNC obtains a higher FoM at lower retained fractions. Neither NNC improves the FoM for the Buzzard sample, but there are a range of retained fractions for which the Misclassification NNC does not hurt the FoM either. See Appendix \ref{buzzard_appendix} for further discussion. Figure \ref{fig:optimal_misclass_sample} shows the sample selection by the Misclassification NNC with the optimized bin edges and retained fraction. The retained fractions achieving the highest FoM in each combination of bin edges and NNC sample selection are listed for CosmoDC2 in Table \ref{table:summary_table}.

\begin{figure*}[ht!]
\plotone{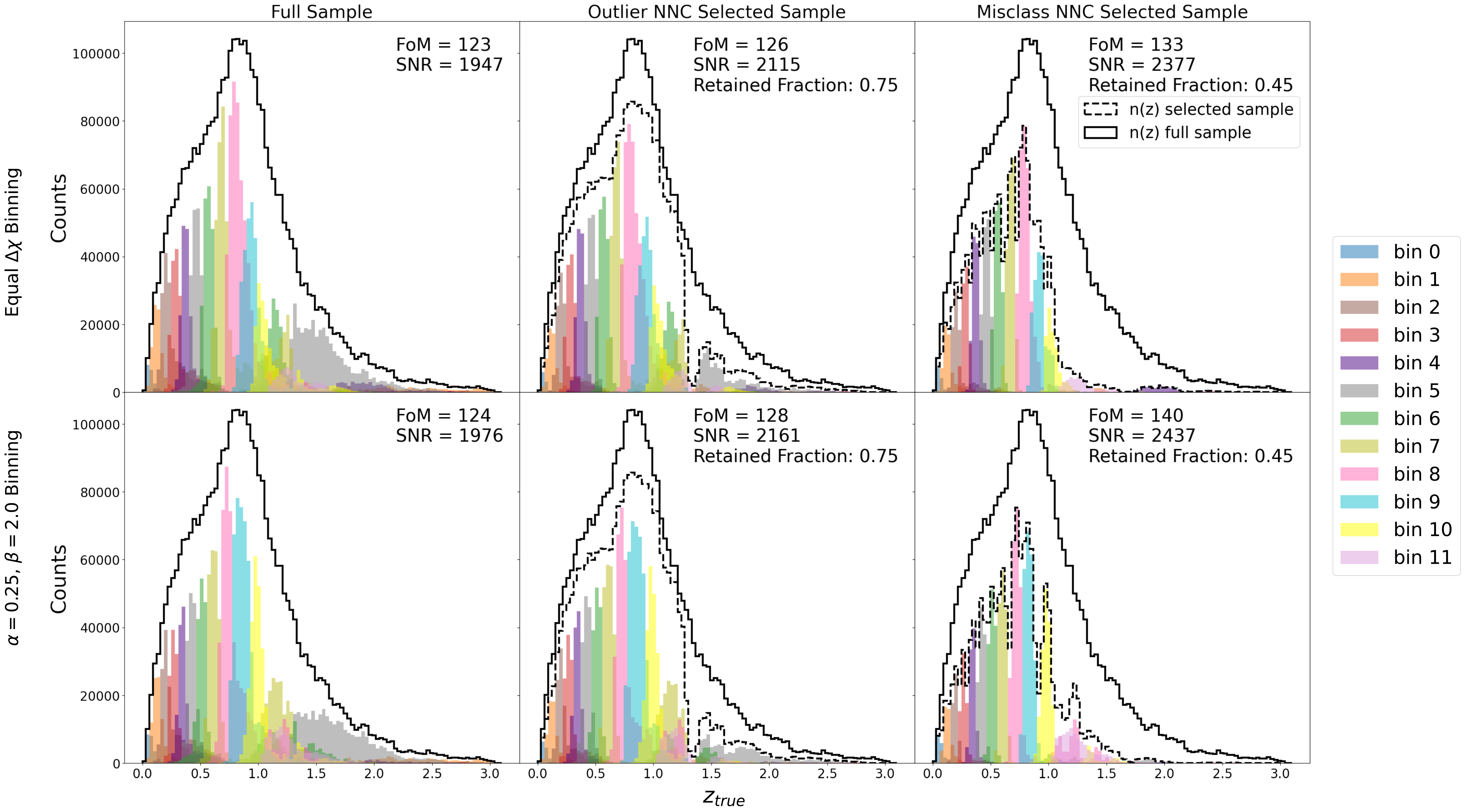}
\caption{A comparison of the NNC performance for each binning method. Top row: equal $\Delta \chi$ binning. Bottom row: optimized $\alpha=0.25$, $\beta=2.0$ binning. The solid black lines show the true $n(z)$ distribution for the full sample, while the dashed lines show the true $n(z)$ distribution for the NNC selected sample in each panel. The retained fraction has been selected to reach the highest possible FoM for each combination of binning type and NNC sample selection. \label{fig:final_bins_cosmodc2}}
\end{figure*}

Both CosmoDC2 and Buzzard obtain a higher FoM with the Misclassification NNC than with the Outlier NNC, suggesting that the Misclassification NNC is preferred.  However, as stated in Appendix \ref{buzzard_appendix}, the highest FoM obtained with the Misclassification NNC is the same as the full sample FoM for Buzzard. Interestingly, although there is not a large improvement in the FoM when the full CosmoDC2 sample of galaxies is sorted into equal $\Delta \chi$ bins versus the optimized bins, the Misclassification NNC greatly increases this improvement at the optimal retained fraction. 

\subsection{Final Binning Results}

\begin{deluxetable*}{cccCCC}

\tabletypesize{\small}
\tablewidth{0pt}
\tablecolumns{6}
\tablecaption{Summary of the bin choices and sample selections. \label{table:summary_table}}

\tablehead{
\colhead{Binning Type} & \colhead{Bin Edges (photo-$z$)} & \colhead{Sample Selection} & \colhead{Retained Fraction} & \colhead{Contamination Fraction} & \colhead{FoM}} 

\startdata
& [0, 0.0758, 0.1548, 0.2372, 0.3234,   & Full & 1 & 0.5370 & 123 \\
Equal $\Delta \chi$ & 0.4142, 0.5102, 0.6120, 0.7206, & Outlier NNC & 0.75 & 0.4374 & 126 \\
& 0.8365, 0.9614, 1.096, 1.242] & Misclassification NNC & 0.45 & 0.1710 & 133 \\
\hline
& [0, 0.0853, 0.1568, 0.2275, 0.3015, & Full & 1 & 0.5310 & 124 \\
Optimized & 0.3792, 0.4634, 0.5550, 0.6557, & Outlier NNC & 0.75 & 0.4261 & 128 \\
& 0.7662, 0.8897, 1.040, 1.242] & Misclassification NNC & 0.45 & 0.1748 & 140 \\
\hline
Representative Sample & - & - & - & - & 186\\
Optimized  & & & & &  
\enddata

\end{deluxetable*}

The Outlier NNC selected sample shown in Figure \ref{fig:outlier_sample}, which retains $75\%$ of the original sample is sorted into equal $\Delta \chi$ bins, which is the best performing of the original three binning choices. The same Outlier NNC selected sample is sorted into the optimized binning choice defined by $\alpha=0.25$ and $\beta=2.0$. The Misclassification NNC selected samples for equal $\Delta \chi$ binning (shown in Figure \ref{fig:misclass_num_sample} with a retained fraction of $75\%$), and for the optimized binning choice, shown in Figure \ref{fig:optimal_misclass_sample}, each with a retained fraction of $45\%$, are sorted into equal $\Delta \chi$ bins and the optimized bins respectively. 

The top row of Figure \ref{fig:final_bins_cosmodc2} shows the binning results for equal $\Delta \chi$ binning, while the bottom row shows the results for the optimized binning choice. In the left column, we have binned the full sample, while the center and right columns show the Outlier NNC and Misclassification NNC selections. Switching from equal $\Delta \chi$ binning to the optimized binning increases the FoM by $\sim 1\%$. The sample selection with the Outlier NNC improves the FoM by a further $3.3\%$, while the Misclassification NNC improves the FoM by $13.2\%$ over using the full sample. The NNC sample selection process boosts the FoMs in the equal $\Delta\chi$ binning case by a similar amount: $3.0\%$ for the Outlier NNC, and $8.5\%$ for the Misclassification NNC. The overall improvement between the Misclassification NNC selected sample sorted into the optimized bins and the full sample sorted into the equal $\Delta \chi$ bins is $14.2\%$. The Misclassification NNC slightly outperforms the Outlier NNC, and in Figure \ref{fig:final_bins_cosmodc2}, there is visibly less overlap between the bins when the Misclassification NNC is used, particularly when compared to the full sample bins. A summary of the different binning methods and sample selections, along with the achieved FoMs, is included in Table \ref{table:summary_table}. 

\begin{figure*}
\plotone{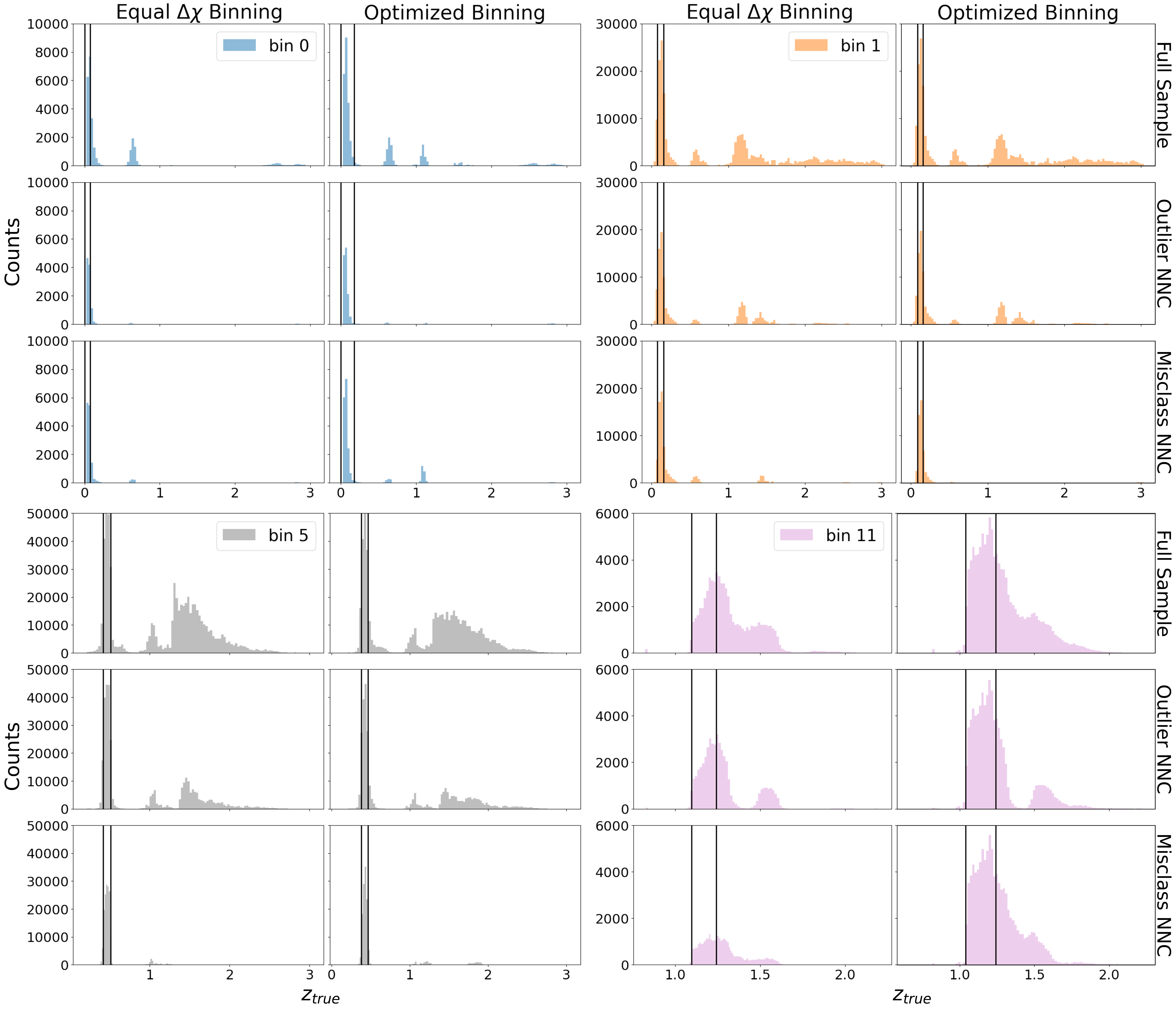}
\caption{A selected set of bins shown individually for each case of bin edges and sample selection. Solid vertical lines indicate the edges of the bins in photo-$z$ space. Upper Left Quadrant: bin 0 contains some contaminating galaxies from higher redshifts. Unlike most other bins, the Outlier NNC is more successful at removing these galaxies. Upper right Quadrant: bin 1 also contains a large number of contaminating galaxies from higher redshifts. The Misclassification NNC is more successful at removing these galaxies than the Outlier NNC, and it is also more successful at removing them in the optimized bin choice than the equal $\Delta \chi$ choice. Lower Left Quadrant: bin 5 contains a large number of contaminating galaxies from high redshifts.  Similar to the bin 1 case, the Misclassification NNC is better able to remove contaminating galaxies outside the bin edges than the Outlier NNC, but this time it is about equally successful in both bin choices. Lower Right Quadrant: The highest redshift bin is much wider in the optimized case than in the equal $\Delta \chi$ case. Similar to the lowest redshift bin, the Outlier NNC performs better than the Misclassification NNC in this bin. \label{fig:individual_bins}}
\end{figure*} 
 
To compare how well the NNC sample selection can recover the ideal case of a representative training sample, we also test the binning method when TPZ and the NNCs have been trained with a fully representative training sample. When the galaxies are sorted into the optimized bin edges, we achieve a FoM of 186, an improvement of $\sim 50\%$ over the non-representative training sample. When the Misclassification NNC is applied in the non-representative case, it is able to recover $\sim 25\%$ of this difference. This indicates that the Misclassification NNC will be useful in attempting to recreate the ideal scenario of a fully representative spectroscopic training sample for LSST.

Given that the Tomographic Challenge already proposed methods for optimizing the tomographic binning strategy for LSST, albeit for the source sample of galaxies instead of the lens sample of galaxies as we have done here, we also compare how the Tomographic Challenge methods compare to the method described in this work. These results are described in Appendix \ref{tomo_challenge_appendix}. 

\subsubsection{Difference Between Binning Choices}

The bin edges for both the optimized bins and equal $\Delta \chi$ bins are listed in Table \ref{table:summary_table}. In the optimized case, the highest redshift bin is $\sim 40\%$ wider than in the equal $\Delta \chi$ case, while most of the rest of the bins are slightly narrower than in the equal $\Delta \chi$ case. This could be a result of the highest redshift bin covering a range that is not well-represented in our training sample, so the photo-$z$ estimates are poorer. We gain constraining power by containing a larger fraction of those galaxies with poor photo-$z$'s in one bin, allowing us to have narrower bins in regions with better photo-$z$'s.

As can be seen in Figures \ref{fig:outlier_sample}, \ref{fig:misclass_num_sample}, and \ref{fig:optimal_misclass_sample}, galaxies at high true redshifts are assigned primarily lower photo-$z$ values. In particular, there are bands at $z\sim 0.1$ and $z\sim 0.5$ where many higher redshift galaxies are assigned. The lower band corresponds to bins 0 and 1, while the $z\sim0.5$ band falls around bin 5. In Figure \ref{fig:individual_bins}, we show a closer look at this subset of the bins to illustrate some differences between the binning choices and sample selections.

In the top rows of the upper left and right quadrants of Figure \ref{fig:individual_bins}, the two lowest redshift bins have significant numbers of contaminating galaxies, where the true redshift places them well outside the bounds of the bin. In the middle and bottom rows of these quadrants, the Outlier and Misclassification NNC are able to remove many of these contaminating galaxies, with the Outlier NNC working better in the lowest redshift bin 0, and the Misclassification NNC doing a better job in bin 1. Additionally, the Misclassification NNC does a slightly better job when applied to the optimized bin 1 than the equal $\Delta \chi$ bin 1.

In the bottom left quadrant of Figure \ref{fig:individual_bins}, we show bin 5, which contains many of the galaxies assigned to the $z\sim0.5$ band. Similar to bin 1, the Misclassification NNC is better able to clean up the contaminating galaxies than the Outlier NNC, although in this case, the Misclassification NNC performs about equally for both choices of bin edges.

In general, the Misclassification NNC removes contaminating galaxies better than the Outlier NNC. The two exceptions are the lowest and highest redshift bins, shown in the upper left and bottom right quadrants of Figure \ref{fig:individual_bins} respectively. In the case of the highest redshift bin, this is likely because not many galaxies are truly misclassified into the highest redshift bin. In general, photo-$z$ estimates are lower than the true redshift, not higher (see, for example, Figure \ref{fig:optimal_misclass_sample}), so there are not a significant number of galaxies misclassified into the highest redshift bin from below. On the other hand, galaxies are not being misclassified from above either, since galaxies assigned into bin 11 that have true redshifts larger than their photo-$z$ would have been assigned to the highest redshift bin anyway. It is not as clear why the lowest redshift bin should also do better with the Outlier NNC as opposed to the Misclassification NNC.

In the remaining bins, the Misclassification NNC performs better at cleaning up contaminating galaxies than the Outlier NNC. In 4 of the 10 remaining bins, the Misclassification NNC cleans up more contaminating galaxies in the optimized binning choice than in the equal $\Delta \chi$ choice. In another 4 bins, it cleans up contaminating galaxies equally well in both bin choices, while in the remaining 2, it cleans up contaminating galaxies better in the equal $\Delta \chi$ case. Given that the Misclassification NNC performs better on the optimized case in a larger percentage of bins, this explains why the optimized bins are able to achieve more constraining power than the equal $\Delta \chi$ bins.

The visual indication that the Misclassification NNC is better able to clean up the contaminating galaxies  is quantified by the contamination fraction. Bin edges are defined in photo-$z$ space; we define the contamination fraction as the fraction of galaxies with $z_{\rm true}$ outside the photo-$z$ range of the bin. When the full sample of galaxies is sorted into equal $\Delta \chi$ bins, the contamination fraction is 0.54; over half the galaxies have $z_{\rm true}$ outside the bin they're sorted into. This is not surprising; the maximum photo-$z$ estimate is $z\sim 1.2$, while the true redshifts extend to $z\sim 3$. With the Outlier NNC selected sample, the contamination fraction is reduced to 0.44. It is further reduced to 0.17 with the Misclassification NNC selection. In contrast, the contamination fraction for the full sample sorted into the optimized bins is 0.53. With the Outlier NNC, this is reduced to 0.43, and to 0.17 with the Misclassification NNC. In both choices of bin edges, the contamination fractions are much lower when the Misclassification NNC is used to select the sample than when the Outlier NNC or no NNC is used. This can be seen visually in the upper right and lower left quadrants of Figure \ref{fig:individual_bins}.

\subsubsection{Impact of NNCs on n(z)}
The NNCs both produce changes in the overall $n(z)$ distribution of the final selected sample compared to the full sample. Since galaxies with true redshifts above $z\sim 1.2$ are not represented in the TPZ training sample, photometric redshifts for those galaxies are primarily outliers. These leads to the majority of galaxies with $z_{\rm true} \gtrsim 1.2$ being removed when the Outlier NNC is used for sample selection. In the middle column of Figure \ref{fig:final_bins_cosmodc2} , we can see that this produces a sharp cutoff in the $n(z)$ distribution above $z\sim 1.2$ in both the equal $\Delta \chi$ and optimized bins.

When the Misclassification NNC is used for sample selection, the $n(z$) distribution is less smooth over the bin edges, and there are notable decreases in the $n(z)$ distribution right at the bin edges. This arises because galaxies with redshifts near the bin edges are more likely to be sorted into the wrong bin, even if the difference between their photo-$z$ estimate and true redshift is small, meaning the Misclassification NNC preferentially removes galaxies near the bin edges, which is not true for the Outlier NNC. In the rightmost column of Figure  \ref{fig:final_bins_cosmodc2}, it can be seen that the $n(z)$ distribution falls off much more gradually above $z\sim 1.2$ than when the Outlier NNC is used, but there are sharp dips around the bin edges.

This modification to the $n(z)$ resulting from the NNCs could introduce biases in the estimated redshift distribution of the sample. Throughout this work, we have assumed that the true $n(z)$ can be perfectly known via spectroscopic cross-correlation, but in practice this will be imperfect. The impact of the binning choices and sample selection with the NNCs on the estimated $n(z)$, as well as any biases in cosmological parameters that may arise from systematic mis-estimation of that distribution, are the subject of future work.

\subsubsection{Combining NNCs}
As seen in Figure \ref{fig:individual_bins}, the Outlier NNC performs better at removing contaminating galaxies in the highest and lowest redshift bins, while the Misclassification NNC is better at removing contaminating galaxies in the intermediate redshift bins. Given this, we test if combining the two NNCs can improve the FoM even further. To do this, we sort the full sample of galaxies into the optimized bin edges, then apply the Outlier NNC to galaxies in the highest and lowest redshift bins, using the optimized retained fraction for the Outlier NNC. We apply the Misclassification NNC, with the appropriate retained fraction, to all other bins. This method achieves a FoM of 105, lower than the FoM achieved while using either NNC alone, or using no NNC at all. Although there is room for attempting to optimize this combination further by adjusting the retained fraction within each bin, we leave this to future work.

\section{Summary and Conclusions}
We have explored two ways to optimize the tomographic binning of the lens sample of galaxies for 3x2pt. analysis: how to optimize the choice of bin edges, and how to optimize the choice of galaxies to sort into those bins. We used HSC DR2 to build a realistically non-representative training sample of galaxies from the CosmoDC2 and Buzzard simulated catalogs. In particular, the training set (with spectroscopic redshifts) is brighter than the application set (with only photometric redshifts), which is expected for LSST data. We then used this sample of galaxies to train TPZ and obtain photometric redshift estimates for the application sample. Galaxies were initially binned by redshift into bins with edges either equally spaced in redshift or comoving distance, or with an equal number of objects in each bin. We introduced the new parameter $\mathcal{M}$ to explore other possible binning choices, and found that although equal $\Delta \chi$ binning performs the best out of the three initial binning choices, it is not the overall optimal choice of bin edges in all cases. For CosmoDC2, the optimized binning choice corresponds to $\alpha=0.25$ and $\beta=2.0$, while Buzzard prefers $\alpha=1.0$ and $\beta=0$, corresponding to equal $\Delta \chi$ binning . This difference in the optimized binning choice indicates
that the optimized binning strategy depends on the underlying $n(z)$ distribution of the galaxies. If this method was to be applied to real data, the optimized bin edges will have to be determined separately for a given galaxy survey, as the choice of $\alpha$ and $\beta$ depends on the underlying $n(z)$ distribution of the galaxy population observed in the survey. However, given that both CosmoDC2 and Buzzard prefer binning choices close to (or exactly) equal $\Delta \chi$ bins, this is a good starting choice. 

After obtaining photo-$z$ estimates, we trained the ``Outlier'' Neural Network Classifier introduced in \citet{Broussard} to estimate the confidence that each photo-$z$ estimate is an outlier, with an NNC confidence near 0 likely to be an outlier. Since galaxies are binned tomographically, and the individual redshift estimates will no longer matter once the binning has been performed, we also trained the NNC in a new way: to estimate the confidence that each photo-$z$ estimate reflects the galaxy being sorted into the correct bin. For this Misclassification NNC, a confidence value near 0 indicates a high probability that the galaxy has been misclassified and sorted into an incorrect redshift bin. We used the NNC confidence values to make sample cuts, selecting only galaxies with high confidence values in each case. The remaining galaxies were then sorted into the same bins defined before.

Figure \ref{fig:retained_frac_dc2} showed that the improvement to the FoM depends on the amount of the sample that is cut. The CosmoDC2 sample reaches the highest FoM when the Misclassification NNC is used to remove the worst $55\%$ of the sample when galaxies are sorted into the optimized binning choice. In general, the Misclassification NNC performs better than the Outlier NNC when larger fractions of the sample are removed, while the Outlier NNC performs better when smaller fractions are removed. Overall, the Misclassification NNC produces the highest FoM in both CosmoDC2 and Buzzard.

One consequence of the NNC sample selection is the modification to the $n(z)$ distribution. With the Outlier NNC, galaxies with true redshift greater than $z\sim1.2$ are mostly removed, leading to a sharp cutoff in the $n(z)$ distribution at redshifts higher than this. When using the Misclassification NNC, galaxies with redshifts near the bin edges are more likely to be sorted into the incorrect bin, so the Misclassification NNC preferentially removes galaxies near the bin edges. This results in an $n(z)$ distribution that is much less smooth than the $n(z)$ for the full sample. Future work will investigate any bias in cosmological parameter estimation resulting from using samples selected with the NNCs, as well as any bias that may arise from the different binning choices.

The methodology developed here for optimizing lens galaxy samples is designed to be broadly applicable, offering independent steps of optimizing the bins and using a Neural Network Classifier to remove galaxies most likely to be photo-$z$ outliers.  
The true optimized choices for a given survey will depend upon the redshift distribution and photo-$z$ probability distributions of its galaxies.  While we have only utilized the TPZ code for photometric redshift estimation, \citet{Broussard} found similar improvements from the application of NNCs for the template-based code BPZ.   For both the CosmoDC2 and Buzzard examples, we found that among the three simplified binning choices, equal $\Delta \chi$ binning leads to higher FoMs, with the opportunity to further improve the FoM by shifting to the more flexible binning arrangement we introduced.  The NNCs offer a significant further reduction in both the outlier and contamination fractions, and the Misclassification NNC can improve the FoM by over $13\%$.  Further investigation of biases in the subsequent measurement of cosmological parameters will be needed to determine if this reduction in outlier fraction yields benefits at that crucial stage of the analysis.

\section*{Acknowledgements}
This paper has undergone internal review by the LSST Dark Energy Science Collaboration. The internal reviewers were Boris Leistedt, Jan Luca van den Busch and Angus Wright. The authors would like to thank Fran\c{c}ois Lanusse for his contribution to the metrics code.

IM, EG, and ABr acknowledge support for this research from the LSST Corporation via grant \#2021-42. IM, EG and ABr also acknowledge support from the U.S. Department of Energy, Office of Science, Office of High Energy Physics Cosmic Frontier Research program under Award Number DE-SC0010008. JN acknowledges support from the U.S. Department of Energy, Office of Science, Office of High Energy Physics Cosmic Frontier Research program under Award Number DE-SC0007914.

Author contributions are as follows. IM updated the NNC framework to include the misclassification training, performed the analysis and wrote the majority of the paper. EG designed the project and supervised the analysis, wrote small sections and provided feedback on the text. ABa wrote the \textsc{FunBins} tomographic binning method. ABr wrote the original Outlier NNC. JN suggested the color-based redshift cut used in creating the non-representative training sample. JZ compiled mock catalogs and contributed to the metrics code.

The DESC acknowledges ongoing support from the Institut National de 
Physique Nucl\'eaire et de Physique des Particules in France; the 
Science \& Technology Facilities Council in the United Kingdom; and the
Department of Energy, the National Science Foundation, and the LSST 
Corporation in the United States.  DESC uses resources of the IN2P3 
Computing Center (CC-IN2P3--Lyon/Villeurbanne - France) funded by the 
Centre National de la Recherche Scientifique; the National Energy 
Research Scientific Computing Center, a DOE Office of Science User 
Facility supported by the Office of Science of the U.S.\ Department of
Energy under Contract No.\ DE-AC02-05CH11231; STFC DiRAC HPC Facilities, 
funded by UK BEIS National E-infrastructure capital grants; and the UK 
particle physics grid, supported by the GridPP Collaboration.  This 
work was performed in part under DOE Contract DE-AC02-76SF00515.

\begin{figure*}
\plotone{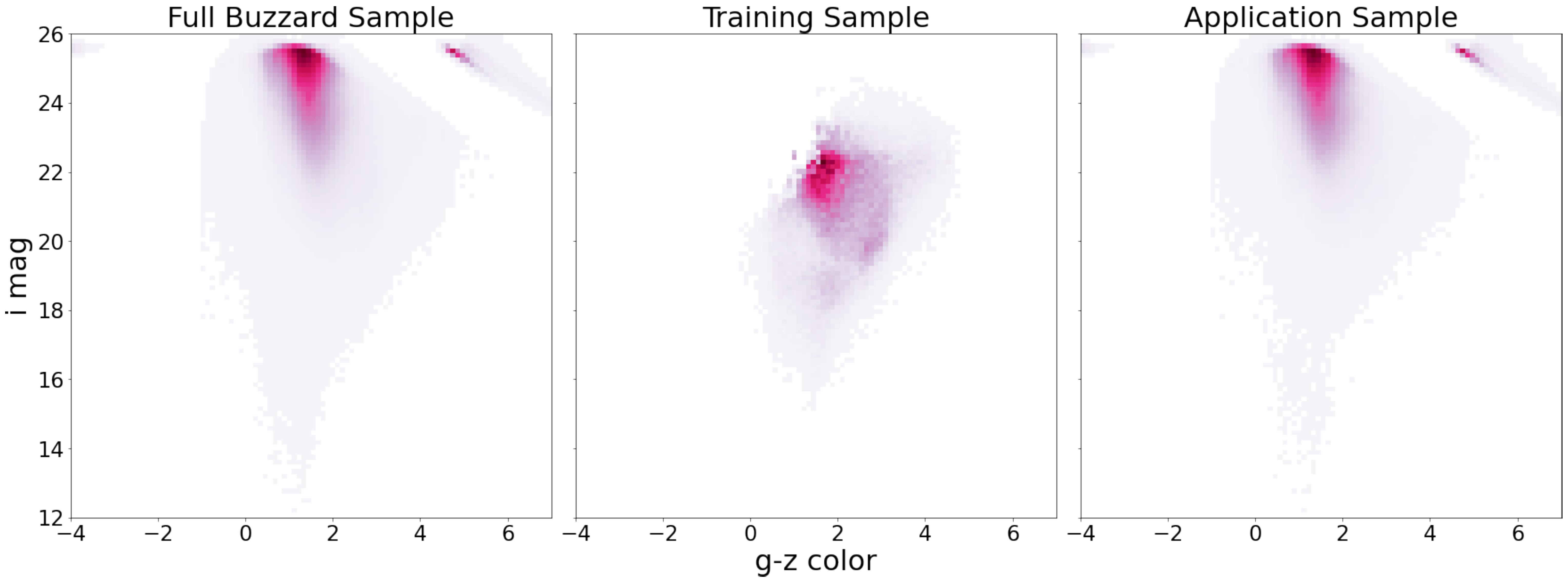}
\caption{Training and application sample selection for Buzzard. The training sample is brighter and slightly redder than the application sample, as is expected for real LSST data.}
\label{fig:unrep_samples_buzzard}
\end{figure*}

\appendix
\section{Buzzard Results}
\label{buzzard_appendix}

\begin{figure}
\plotone{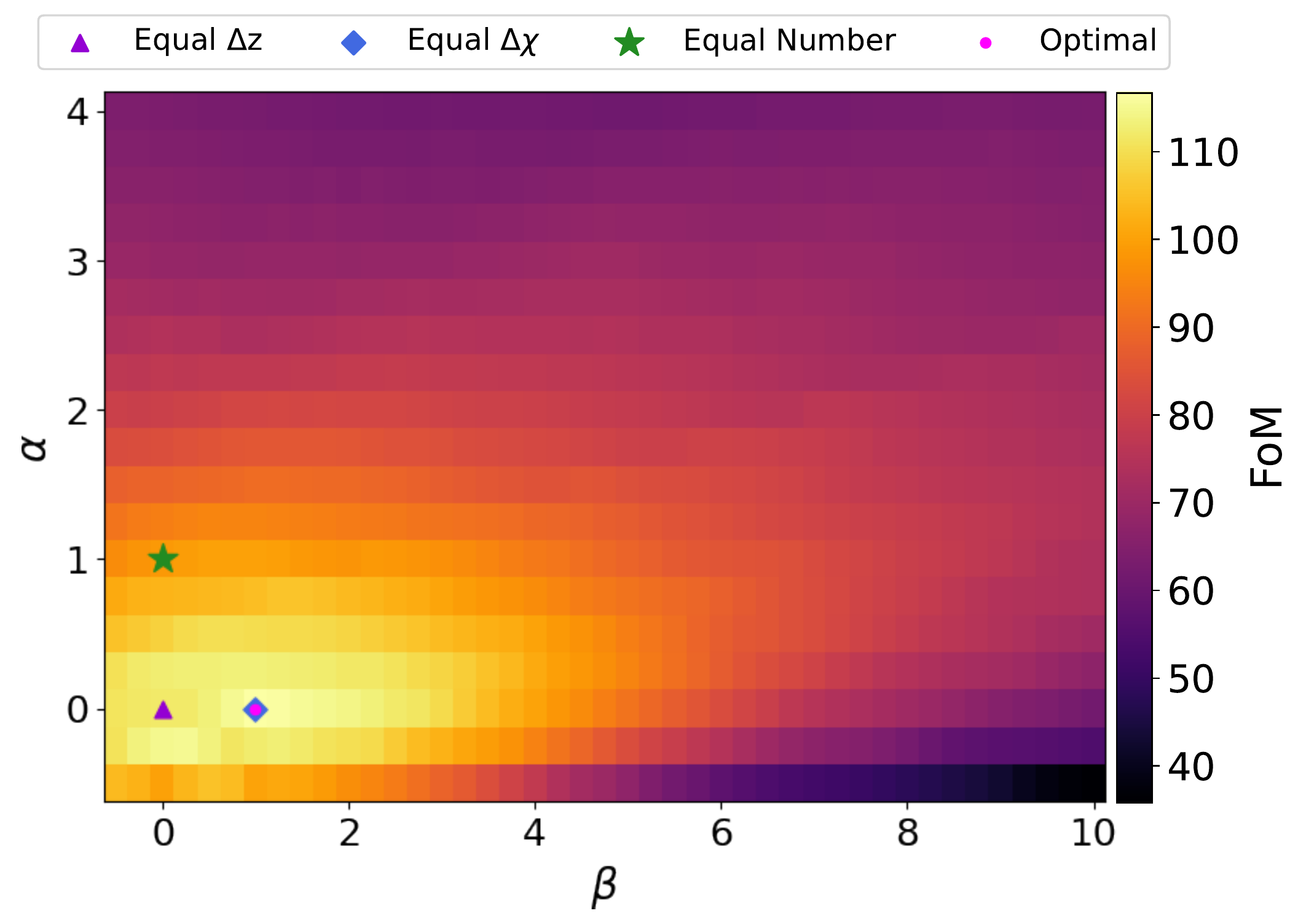}
\caption{The FoMs of the resulting bins vs different combinations of $\alpha$ and $\beta$ for the Buzzard sample. The maximum FoM of 44 is achieved at $\alpha=1.0$, $\beta=0.0$, corresponding to equal $\Delta \chi$ binning. The optimal combination of $\alpha$ and $\beta$ is shown as the pink dot, while the previous binning choices of equal $\Delta z$, equal $\Delta \chi$ and equal number binning are shown as a purple triangle, blue diamond and green star, respectively. Note that the symbols for the optimal binning and equal $\Delta \chi$ binning overlap.}
\label{fig:alpha_beta_buzzard}
\end{figure}

\begin{figure}
\plotone{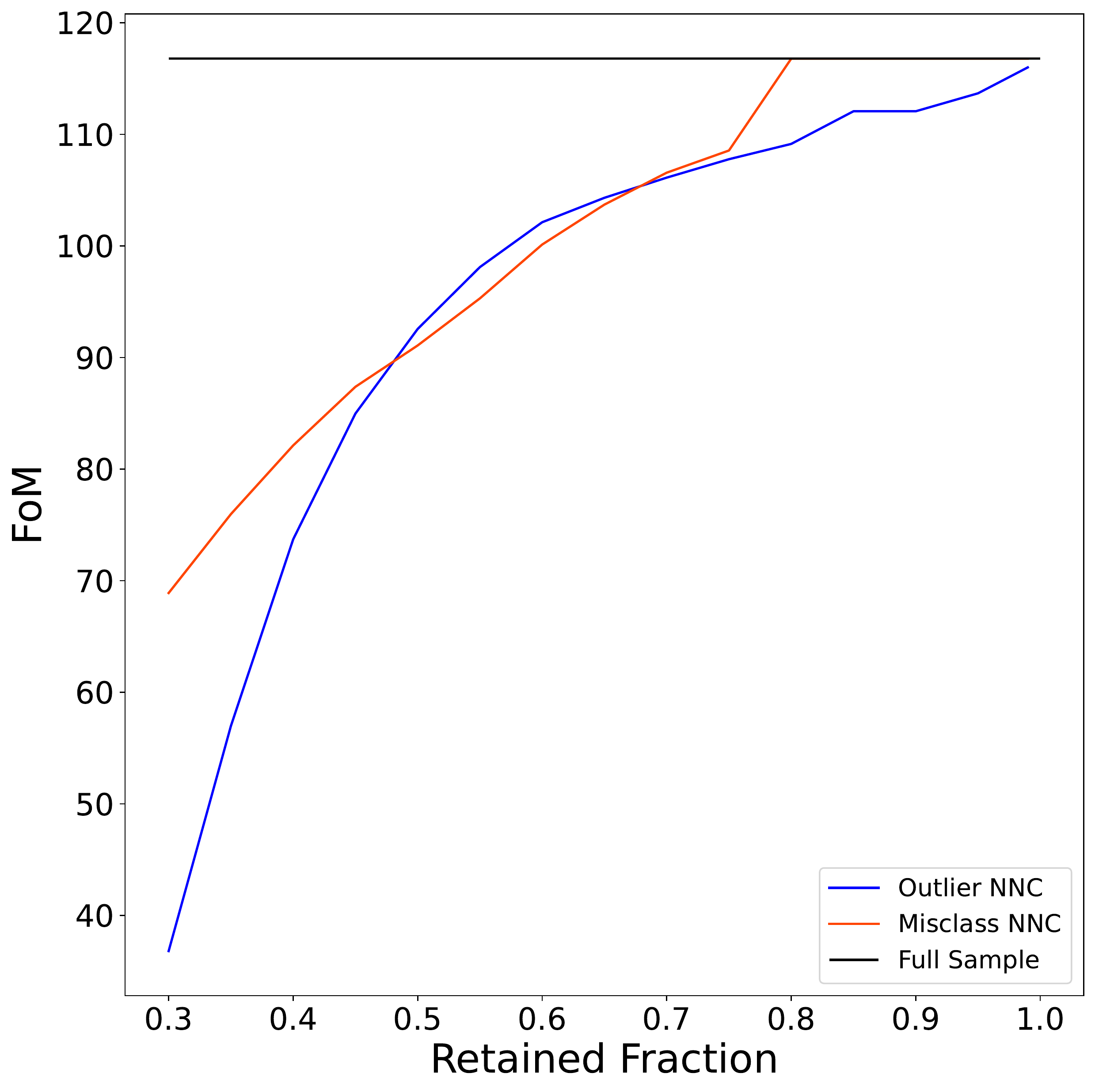}
\caption{Retained fractions vs. FoM for the Buzzard sample sorted into equal $\Delta \chi$ bins, and with samples selected by the Outlier NNC (blue line) and the Misclassification NNC (red line). The horizontal black line shows the FoM for the full sample as a reference. Unlike CosmoDC2, neither NNC improves the FoM for the Buzzard sample.}
\label{fig:retained_frac_buzzard}
\end{figure}

We use the same process to divide the Buzzard data set into non-representative training and application samples. The resulting samples are shown in Figure \ref{fig:unrep_samples_buzzard}. Similar to the partitioning of the CosmoDC2 sample, the resulting training sample has a median \textit{i}-band magnitude of 21.3, brighter than the median \textit{i}-band magnitude of the application sample at 24.3. The training sample is again redder, with a median (\textit{g-z}) color of 2.00 compared to 1.51 for the application sample.

We determine the optimal values of $\alpha$ and $\beta$ for the choice of bin edges using the true $n(z)$, as we did for CosmoDC2. The results are shown in Figure \ref{fig:alpha_beta_buzzard}. Similar to the CosmoDC2 sample, lower values of $\alpha$ and $\beta$ are preferred. However, where CosmoDC2 prefered $\alpha$ and $\beta$ close to, but not exactly, equal $\Delta \chi$ binning, Buzzard attains the maximum FoM at $\alpha = 0$, $\beta=1.0$, corresponding to equal $\Delta \chi$ binning. This difference is mostly likely due to the difference in $n(z)$ distributions between the CosmoDC2 and Buzzard simulations. In particular, the Buzzard sample has a lower maximum redshift than CosmoDC2. The optimal values of $\alpha$ and $\beta$ are not universal, but both simulations agree that equal $\Delta \chi$ binning is at least a close approximation of the optimal binning choice.

We also investigated the optimal retained fraction when using the Outlier and Misclassification NNCs with equal $\Delta \chi$ bins. Figure \ref{fig:retained_frac_buzzard} shows the results. Unlike CosmoDC2, neither NNC improves the FoM for the Buzzard sample. Although the FoM drops for all retained fractions when the Outlier NNC is used, when the Misclassification NNC is applied, retained fractions above 0.75 still attain the same FoM as the full sample. Similarly to CosmoDC2, the Misclassification NNC achieves higher FoM than the Outlier NNC, so although the Misclassification NNC does not improve the FoM overall, it still performs better than the Outlier NNC. The difference in NNC behavior between CosmoDC2 and Buzzard further shows that the binning optimization process must be repeated for different surveys with different $n(z)$ distributions. 

\section{Comparison to Tomographic Challenge Results}
\label{tomo_challenge_appendix}

The Tomographic Challenge was conducted to determine a method of optimizing the source sample binning, as opposed to the lens sample binning we have optimized in this work. The winning method of the Tomographic Challenge was \textsc{FunBins} \citep{tomo_challenge}, which uses bin edges equally spaced in comoving distance $\chi$. Galaxies are assigned to bins using a random forest classifier trained with some combination of colors and magnitudes, but they are not assigned specific photo-$z$ values. 

\begin{figure}
\plotone{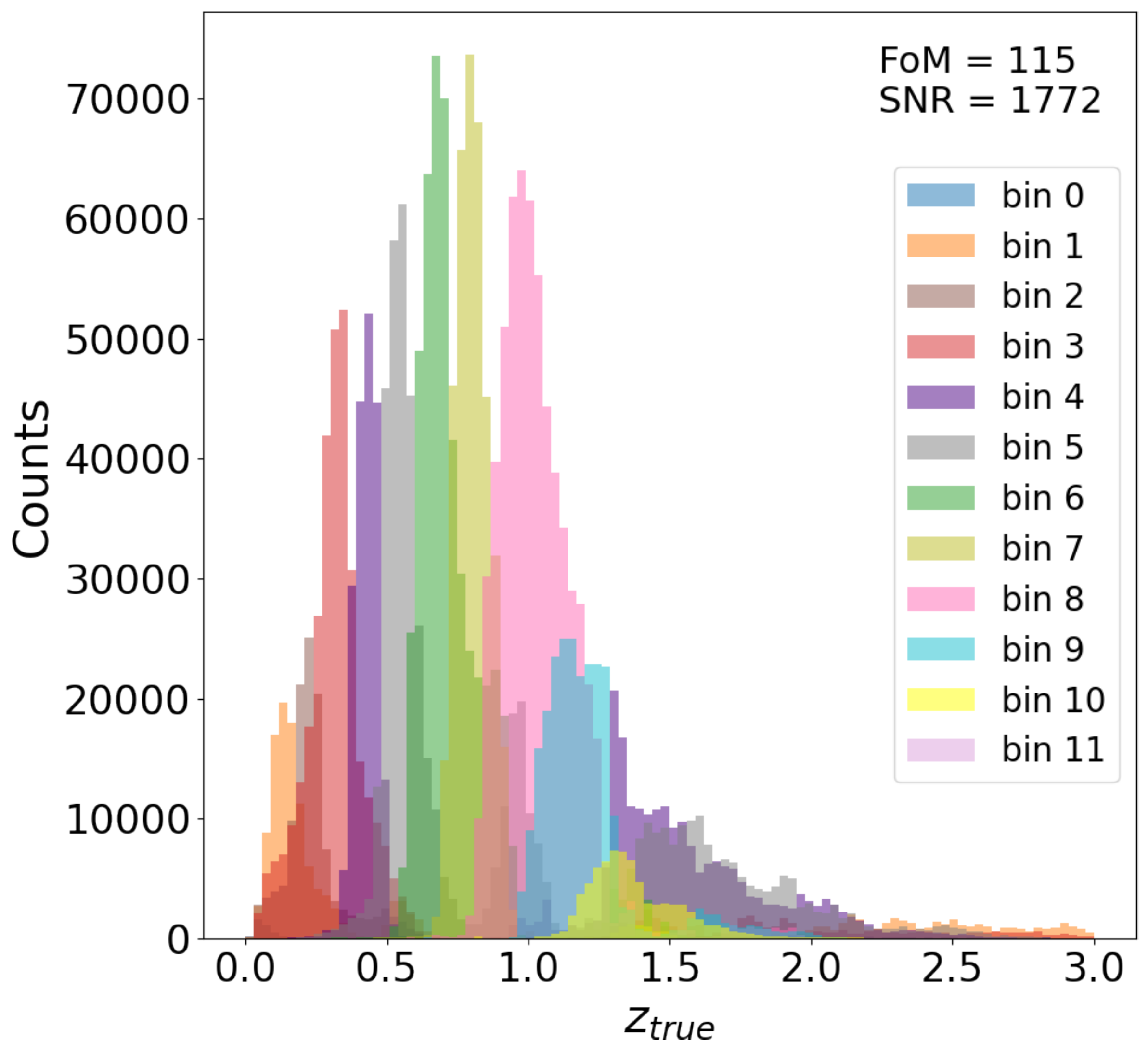}
\caption{Binning results when the \textsc{FunBins} Tomographic Challenge method is trained using the CosmoDC2 training set, colors and magnitudes as described in this work. The FoM is lower than can be achieved with the methods described in this work (see Figure \ref{fig:final_bins_cosmodc2} and Table \ref{table:summary_table}). \label{fig:funbins}}
\end{figure}

Although the Tomographic Challenge was conducted with a fully representative training set that was large compared to the application set, we can train \textsc{FunBins} using our non-representative training set to compare the method outlined in this paper with the results of the Tomographic Challenge. We train \textsc{FunBins} using the same colors and magnitudes as were provided to TPZ for training. The results produced by \textsc{FunBins} using this training set for CosmoDC2 are shown in Figure \ref{fig:funbins}. \textsc{FunBins} is able to achieve a FoM $\sim 115$ under these training conditions, equivalent to our equal $\Delta$z bins and a $6.5\%$ decrease from our full sample sorted into equal $\Delta \chi$ bins.Compared to our NNC selected samples, the \textsc{FunBins} FoM is a $14\%$ decrease from the maximum FoM achieved with equal $\Delta \chi$ bins, and an $18\%$ decrease from the maximum FoM achieved with our optimized bins for CosmoDC2 (see Table \ref{table:summary_table}). To achieve the same FoM as our optimized tomography and NNC selection method, a more conventional method, such as \textsc{FunBins}, would be required to observe an additional $\sim18\%$ of the total LSST Year 1 area.

\bibliographystyle{aasjournal}
\bibliography{biblio}

\end{document}